\documentclass[article,reqno]{amsart}
\usepackage{amsmath,amssymb,mathtools}
\usepackage{latexsym}
\usepackage{epsfig}
\usepackage{booktabs}
\usepackage{color}
\usepackage{graphicx}
\usepackage{tikz}
\usepackage{array,booktabs}
\usepackage{pgfplots}
\usepackage{physics}
\newcommand{\sign}{\text{sign}}

\newcommand{\beq}{\begin{equation}}
\newcommand{\eeq}{\end{equation}}
\newcommand{\beqnl}{\begin{eqnarray}}
\newcommand{\eeqnl}{\end{eqnarray}}

\newcommand{\be}{\begin{equation}}
\newcommand{\ee}{\end{equation}}
\newcommand{\bes}{\begin{equation*}}
\newcommand{\ees}{\end{equation*}}
\newcolumntype{C}{>{\centering\arraybackslash}m{1in}} 

\numberwithin{equation}{section}

\title{Analogue Hawking Effect: a master equation}
\author{F. Belgiorno$^{1,2,3}$, S.L. Cacciatori$^{2,4}$ and A. Vigan\`o$^{2,5}$}
\address{\noindent $^1$Dipartimento di Matematica, Politecnico di Milano, Piazza Leonardo 32, IT-20133 Milano, Italy\endgraf
$^2$INFN sezione di Milano, via Celoria 16, IT-20133 Milano, Italy\endgraf
$^3$INdAM-GNFM \endgraf
$^4$Department of Science and High Technology, Universit\`a dell'Insubria, Via Valleggio 11, IT-22100 Como, Italy\endgraf
$^5$Dipartimento di Fisica, Universit\`a degli Studi di Milano, Via Celoria 16, IT-20133 Milano, Italy\endgraf
}

\begin{document}

\begin{abstract}
We consider further on the problem of the analogue Hawking radiation. We 
propose a fourth order ordinary differential equation, which allows to discuss the 
problem of Hawking radiation in analogue gravity in a unified way, encompassing fluids 
and dielectric media. In a suitable approximation, involving weak dispersive effects, WKB solutions are obtained far from the horizon (turning point), 
and furthermore an equation governing the behaviour near the horizon is derived, and a complete set 
of analytical solutions is obtained also near the horizon. The subluminal case of the original fluid model introduced by Corley and Jacobson, the case of
dielectric media are discussed. We show that in this approximation scheme there is a mode which is not 
directly involved in the pair-creation process. Thermality is verified and a framework for calculating the grey-body factor is provided.  
\end{abstract}

\maketitle

\section{Introduction}

The analogue Hawking effect has been largely discussed in literature, and we are interested to focus our attention on the analytical side of calculations in presence of dispersion. As is well known, the 
problem is very hard and requires techniques borrowed from asymptotic analysis, see e.g. the following (non-exhaustive) list of papers~\cite{brout,corley,himemoto,saida,Schutzhold-Unruh,balbinot,unruh-s,cpf,Leonhardt-Robertson,Coutant-prd,Coutant-und,Un-schu,petev,Coutant-thick,hopfield-hawking,linder,philbin-exact,hopfield-kerr,coutant-subcritical,coutant-kdv,coutant-bdg}. Even if the 
mathematics to be adopted is quite similar, still different systems seem to require different tools 
to be discussed, and what is done for fluids is not just the same as for dielectric media. 
Even if a strictly unified framework a priori is not mandatory, still it is interesting to point out that such a framework exists and allows to draw common conclusions for the various physical situations at hand, and to 
realize an universality for the analogous Hawking effect (see e.g.~\cite{unruh-s}).

In this paper, we propose a fourth order ordinary differential equation as a master equation allowing 
to deal with the analogous Hawking effect in condensed matter systems in a systematic way, in the 
approximation of weak dispersive effects.  
This is {\sl per se} interesting, because (1) a single master equation is shown to be enough for describing different physical situations. In this paper we deal with the subluminal version of the fluid model introduced by Corley and Jacobson \cite{corley-jacobson,corley}, and also with the case of dielectric media. In the 
companion paper~\cite{belbecsw}
we discuss also the case of the analogous Hawking effect in BEC (superluminal case), and in water.

As a second element of interest, (2) a single approximation is done, allowing to reduce the problem into a form which is analogous to the one described in a series of works by~\cite{la,ni,ni-I,ni-global,ni-tp}. {\color{black} To be more specific, 
we adopt the limit of weak dispersive effects in all models (for the previous literature, concerning analytical calculations, see  e.g. in \cite{cpf,Coutant-thick}.)} 
Furthermore, (3) a new kind of near-horizon expansion (expansion near the turning point) is adopted, allowing to get a completeness of states also in that physical region{\color{black}, in particular we can take into account explicitly the -$s$-modes (also called $v$-modes; see section~\ref{corleywkb}) which are neglected in other near-horizon expansions;} {\color{black} furthermore, the near-horizon equation one obtains is universal, i.e. it has the same form for all the models we take into account, and this is at the root of the universality of the Hawking effect in analogue gravity;}
(4) the nature of the horizon (turning point) is clearly emerging, and the role 
of both $v/c$ in the fluid models, and of the horizon equation $n =c/v$ (phase horizon) in the dielectric case are enhanced. 
Connection formulas allow to 
calculate  the fundamental ratio $|J_x^+|^2/|J_x^-|^2$, where $J_x^\pm$ stays for the (conserved) current 
associated with the dispersive modes of wavenumbers $k_\pm$ ($k_-$ is associated with negative norm) (see 
sections~\ref{corleywkb},~\ref{corleynhe}). As well known, this ratio qualifies thermality of the Hawking 
analogue radiation.
Last, but non least, (5)  
one may also provide a general rule for the computation the grey-body factor, {\color{black} which is in agreement with the analysis 
carried out in \cite{cpf,Coutant-thick} as far as the Corley model is concerned, and that is extended to the dielectric model to be discussed herein.} 
As general assumptions, in agreement with the aforementioned previous literature, we consider the situation where dispersive effects are mild and the relevant background fields like $v(x)$, $c(x)$ in the fluid models and $n(x)$ in the dielectric models are asymptotically constant and bounded. 
In a remarkable correspondence with the standard black hole case, the grey-body factor is simply due to `scattering on a barrier', provided by the geometry, of the Hawking modes created in the region of the horizon, and is not directly 
associated with the presence of the horizon itself. The fourth wavenumber mode, a short wavenumber mode distinct from the Hawking mode, is then actually decoupled at the horizon. 

\section{The master equation: a Orr--Sommerfeld type fourth order equation}

We show that three significant cases of wave equations in dispersive analogue gravity can be 
reconduced to the equation
\beq
\label{orso}
\epsilon^2 \frac{d^4 \Phi}{dx^4}\pm\biggl[ p_3 (x,\epsilon) \frac{d^2 \Phi}{dx^2}+
p_2 (x,\epsilon) \frac{d \Phi}{dx} +p_1 (x,\epsilon) \Phi \biggr]=0, 
\eeq
where the upper sign occurs in the case of subluminal dispersion and the lower one in the 
case of superluminal dispersion.
The latter case is considered in Nishimoto's
works (see e.g.~\cite{ni} and references therein). Furthermore, 
\beq
p_i (x,\epsilon) = \sum_{n=0}^\infty p_{i n} (x) \epsilon^n ,
\eeq
is assumed.
As $\epsilon \to 0$ one finds the so-called reduced equation
\beq
p_{30} (x) \frac{d^2 \Phi}{dx^2}+
p_{20} (x) \frac{d \Phi}{dx} +p_{10} (x) \Phi =0 .
\eeq
Solutions of 
\beq
p_{30}(x)=0
\eeq
define the turning points (TPs) of the equation, and in the analysis of the reduced equation 
the behaviour of solutions in the neighbourhood of the TPs is of utmost relevance for the 
scattering problem we mean to delve into.
In the following, we limit ourselves to the case of a single TP, to be identified with $x=0$ without loss of generality. 
In~\cite{ni} it is assumed that the reduced 
equation displays a Fuchsian singularity at the TP (nothing actually prevents the 
general equation in itself to admit a regular behaviour). One may then expect two kinds of 
solutions:
\begin{align}
\Phi^{(1)} & = 1 + \sum_{n=1}^\infty d_n x^n , \\
\Phi^{(2)} & = x^{1-\lambda} \biggl(1+ \sum_{n=1}^\infty e_n x^n\biggr) ,
\end{align}
where $\lambda$ is related to a root of the so-called indicial equation associated with the 
reduced equation in the neighbourhood of the TP.
This kind of solution appears to be useful in the 
WKB approximation, which in our scheme, differently from the hypotheses in~\cite{la,ni}, can be extended to hold also in the 
asymptotic region of unboundedly large values of $x$. It is worth mentioning that the first solution above is regular at the 
turning point. This is relevant also in the following sections.

The great advantage of referring to the above equation is that sophisticated analytical calculations 
carried out mostly by~\cite{ni} are just available,
where a considerable effort has to be exploited in order to keep under control the asymptotic formulas and the associated connection formulas. 

\section{A summary of the approximation method near the turning point}

\begin{figure}[hbtp!]
\includegraphics[scale=0.55]{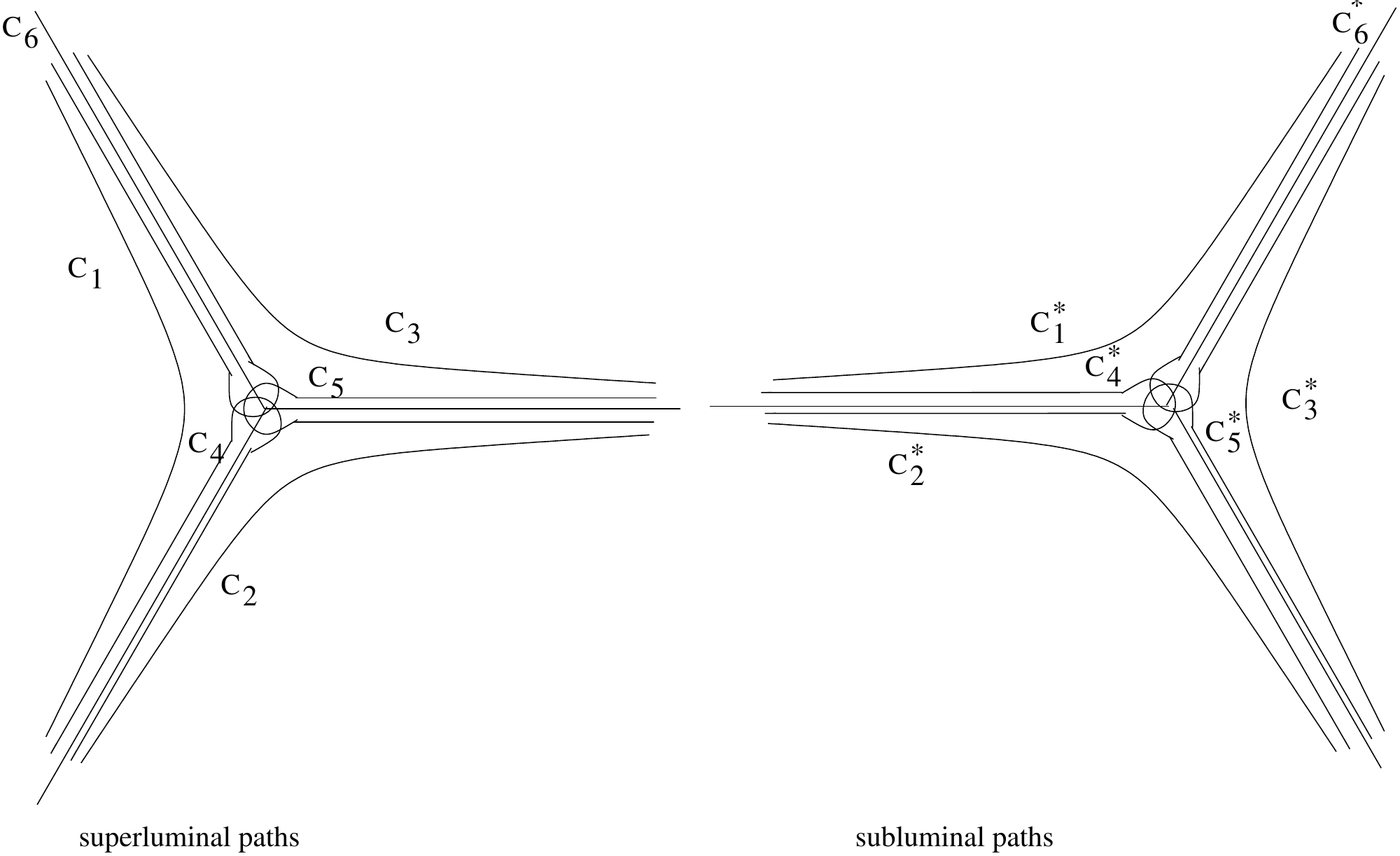} 
\caption{Paths for the superluminal case (left side, see~\cite{ni,ni-global}), labeled with $C_j$, $j=1,\ldots,6$, and 
also for the subluminal case (obtained by a rotation of $-\pi/3$). An asterisk has been introduced for the paths in the latter case.}
\label{fig:paths-ni}
\end{figure}

We sketch for the sake of completeness the essentials of the approximation method near the TP as described in~\cite{ni}, 
of which we maintain the same notation.
The starting point consists in rewriting equation~\eqref{orso} as the first order system
\beq
\label{first}
\epsilon \frac{dY}{dx} = P(x,\epsilon) Y,
\eeq
where 
\beq
Y=\begin{pmatrix}
y \\
y' \\
y'' \\
\epsilon y^{(3)}
\end{pmatrix} , 
\eeq
and 
\beq
P(x,\epsilon) = \begin{pmatrix}
0 & \epsilon & 0 & 0 \\
0 & 0 & \epsilon & 0 \\
0 & 0 & 0 & 1 \\
\mp p_1(x,\epsilon) & \mp p_2(x,\epsilon) & \mp p_3 (x,\epsilon) & 0
\end{pmatrix},
\eeq
where, again, the upper sign is relative to the subluminal case. 
The `stretching and shearing transformations' 
\begin{align}
x-a & = \epsilon^{2/3} s , \\
Y & = \Omega (\epsilon) W , \label{yw}\\
\Omega (\epsilon) & \coloneqq \mathrm{diag} \bigl\{  \epsilon^{4/3},  \epsilon^{2/3} , 1,  \epsilon^{1/3} \bigr\},
\end{align}
where $a$ is the turning point, allow to obtain
\beq
\label{first-tp}
\frac{dW}{ds} = A(s,\epsilon) W,
\eeq
where 
\beq
A(s,\epsilon) = \begin{pmatrix}
0 & 1 & 0 & 0 \\
0 & 0 & 1 & 0 \\
0 & 0 & 0 & 1\\
\mp p_1(x(s),\epsilon) \epsilon^{2/3} & \mp p_2(x(s),\epsilon) & \mp p_3 (x(s),\epsilon) \epsilon^{-2/3} & 0
\end{pmatrix} ,
\eeq
and $x(s)= a + \epsilon^{2/3} s$. 
The functions $p_i(x,\epsilon)$ ($i=1,2,3$) can be expanded in power series of $\epsilon$ with coefficients which are polynomials of 
$x-a$ in the neighborhood of the TP, and in turn the matrix $A$ can be expanded in power series of $\epsilon^{1/3}$ with polynomial coefficients of $s$. 
Solutions are constructed in the form 
\beq
W(s,\epsilon) = \sum_{i=0}^\infty W_i (s) \epsilon^{i/3} ;
\eeq
at the lowest order, $W_0(s)$ must satisfy
\beq
\label{eq-w}
\frac{dW_0}{ds}= A_0 (s) W_0,
\eeq
where 
\beq
A_0 (s) = \begin{pmatrix}
0 & 1 & 0 & 0 \\
0 & 0 & 1 & 0 \\
0 & 0 & 0 & 1\\
0 & \mp p_{20} (a) & \mp p'_{30}  (a) s & 0
\end{pmatrix} .
\eeq
Equation~\eqref{eq-w} is equivalent to the fourth order differential equation 
\beq
\label{four-z}
\frac{d^4 w}{dz^4}\pm \biggl( z \frac{d^2 w}{dz^2} +\lambda \frac{dw}{dz} \biggr)=0,
\eeq
where 
\beq
\label{change-var}
z= (p'_{30}(a))^{1/3} s =(p'_{30}(a))^{1/3} \epsilon^{-2/3} (x-a),
\eeq
and 
\beq
\lambda = \frac{p_{20}(a) }{p'_{30} (a)}.
\eeq
A further corroboration of equation~\eqref{four-z} is contained in Appendix A. 
Solutions to equation~\eqref{four-z} are found by means of Laplace integrals
\beq
\label{gen-airy}
w_j(z) = \frac{1}{2\pi i} \int_{C_j} dt\ t^{\lambda -2} \exp (z t\pm \frac{1}{3} t^3),
\eeq
with a suitable choice for the paths $C_j$ in the complex $t$-plane. 
See e.g.~\cite{ni-global,ni} for the superluminal case, where solutions of~\eqref{gen-airy} are also known as generalized Airy 
functions. See figure~\ref{fig:paths-ni} on the left side for paths $C_j$ adopted in~\cite{ni,ni-global}, 
with $j=1,\ldots,6$.

It is interesting to deduce the solutions above directly, in order to point out the subtleties in solving~\eqref{four-z}. 
We first deduce solutions~\eqref{gen-airy}, by means of the Laplace-transform formalism: by putting
\beq
w_j(z) =  \frac{1}{2\pi i} \int_{C_j} dt\ \phi(t)  \exp (z t), 
\eeq
we find 
\beq
\frac{1}{2\pi i} \int_{C_j} dt\ \bigl( t^4 +z t^2 +\lambda t\bigr) \phi (t)  \exp (z t)=0, 
\eeq
and, as usual, thanks to an integration by parts
\beq
\frac{1}{2\pi i} \int_{C_j} dt\ z t^2 \phi (t)  \exp (z t)=\frac{1}{2\pi i} t^2 \phi (t) \exp (z t)\bigr|_{C_j} - 
\frac{1}{2\pi i} \int_{C_j} dt\ \biggl[ \frac{d}{dt} \bigl(t^2 \phi (t)\bigr)\biggr]  \exp (z t), 
\eeq
where $t^2 \phi (t) \exp (z t)\bigr|_{C_j}$ is the variation of $t^2 \phi (t)$ along $C_j$,  
one obtains solutions by putting 
\beq
\frac{d}{dt} \bigl(t^2 \phi (t)\bigr) =\pm \biggl(t^2 +\frac{\lambda}{t}\biggr) t^2 \phi (t), 
\eeq
which provides us~\eqref{gen-airy}, and imposing also  
\beq
t^{\lambda} \exp (z t\pm \frac{1}{3} t^3)\biggr|_{C_j}=0.
\eeq
The fourth solution, i.e.~the constant solution, which is present in a trivial way as a solution of 
the original equation~\eqref{four-z}, seems to be quite `hidden' in the Laplace-transform formalism. Naively, 
it would seem that one could find it by a suitable choice of the path $C$ along which the complex integration is performed. 
For example, one might easily find the zero-solution as a integral along any closed path non-intersecting the cut. Still, 
this reasoning is too naive. What is a bit hard to realize, is that the equation obtained by Laplace transform admits 
also a distributional solution: indeed, it can be rewritten as 
\beq
t^2  \frac{d \phi(t)}{dt} + \bigl(2 t \mp (t^4+\lambda t)\bigr) \phi(t)=0.
\label{eq-lapl}
\eeq
As a consequence, it is easy to show that 
\beq
\phi (t) = \delta (t) ,
\label{delta-sol}
\eeq
is a distributional solution, where $\delta (t)$ is the Dirac delta. By direct substitution, we get a first term which is 
$t^2 \delta' (t)$, which is zero (cf.~\cite{kanwal}, equation (4), p.~256). At the same time, in a distributional sense, 
we get also $(2 t \mp (t^4+\lambda t)) \delta(t)=0$. Then, the constant solution arises in this framework as
\beq
w_C (z) = \int_C dt\ \delta(t)  \exp (z t)=1.
\label{lapl-const}
\eeq
\\
Note that we are allowed to put (cf.~\cite{cpf})
\beq
z=\sign (z) |z|,
\eeq
as we are interested in real values of $z$ (and $x$).

Paths extending to infinity in the complex $t$-plane must be restricted to allowed regions. In the 
superluminal case ($-$ sign in front of the cubic term in the exponential) we have the same regions as for the 
Airy functions, with $\theta\coloneqq\arg(t)$:
\beq
\theta\in \left(-\frac{\pi}{6},\frac{\pi}{6}\right) \cup \left(\frac{\pi}{2},\frac{5 \pi}{6}\right) \cup \left(\frac{7\pi}{6}, \frac{3\pi}{2}\right).
\eeq
In the subluminal case ($+$ sign in front of the cubic term in the exponential) we find the complementary 
regions:
\beq
\theta\in \left(\frac{\pi}{6},\frac{\pi}{2}\right) \cup \left(\frac{5\pi}{6},\frac{7 \pi}{6}\right) \cup \left(\frac{3\pi}{2}, \frac{11\pi}{6}\right)
\eeq
It is interesting to point out that one may select a basis of solutions. E.g., for the superluminal case, 
we list the approximations of the solutions in the asymptotic region (large $z$) as determined in~\cite{ni}:
\begin{subequations}
\begin{align}
w_1(z) &= -\frac{e^{\lambda \pi i}}{2 \sqrt{\pi}} z^{\frac{\lambda}{2}-\frac{5}{4}} e^{-\frac{2}{3} z^{\frac{3}{2}}} (1+O (z^{-\frac{3}{2}} )), \quad \quad |\arg(z)|<\pi,\\
w_2(z) &= \frac{e^{-\lambda \pi i}}{2 \sqrt{\pi}} z^{\frac{\lambda}{2}-\frac{5}{4}} e^{-\frac{2}{3} z^{\frac{3}{2}}} (1+O (z^{-\frac{3}{2}} )), \quad \quad 
\frac{\pi}{3}<\arg(z)<\frac{7\pi}{3},\\
w_3(z) &= \frac{i}{2 \sqrt{\pi}} z^{\frac{\lambda}{2}-\frac{5}{4}} e^{\frac{2}{3} z^{\frac{3}{2}}} (1+O (z^{-\frac{3}{2}} )), \quad \quad 
-\frac{\pi}{3}<\arg(z)<\frac{5\pi}{3},\\
w_4(z) &= \frac{e^{\lambda \pi i}-e^{-\lambda \pi i}}{2 \pi i} \Gamma (\lambda-1) z^{1-\lambda} (1+O (z^{-3} )), \quad \quad 
-\pi<\arg(z)<\frac{\pi}{3},\\
w_5(z) &= \frac{e^{\lambda \pi i}-e^{-\lambda \pi i}}{2 \pi i} \Gamma (\lambda-1) z^{1-\lambda} (1+O (z^{-3} )), \quad \quad 
\frac{\pi}{3}<\arg(z)<\frac{5\pi}{3},\\
w_6(z) &= \frac{e^{\lambda \pi i}-e^{-\lambda \pi i}}{2 \pi i} \Gamma (\lambda-1) z^{1-\lambda} (1+O (z^{-3} )), \quad \quad 
-\frac{\pi}{3}<\arg(z)<\pi.
\end{align}
\end{subequations}
A basis of solutions is obtained by considering one of the following sets: 
\begin{subequations}
\begin{align}
W_0^{(1)} & \coloneqq \{1,w_6 (z),w_3 (z), w_1(z)\} , \\
W_0^{(2)} & \coloneqq \{1,w_5 (z),w_2 (z), w_3(z)\} , \\
W_0^{(3)} & \coloneqq \{1,w_4 (z),w_1 (z), w_2(z)\},
\end{align}
\end{subequations}
for 
\begin{subequations}
\begin{align}
\arg (z) & \in \left(-\frac{\pi}{3},{\pi}\right) , \\
\arg (z) & \in \left(\frac{\pi}{3},\frac{5\pi}{3}\right) , \\
\arg (z) & \in \left(-{\pi},\frac{\pi}{3}\right) ,
\end{align}
\end{subequations}
respectively~\cite{ni-global}.

It is also easy to show that, both in the superluminal and in the subluminal case, by choosing 
suitably also the subluminal solutions, one finds
\begin{subequations}
\begin{gather}
w_1 (z) = \psi^{\lambda-1} w_3 (\psi z) = \psi^{2(\lambda-1)} w_2 (\psi^2 z),\\
w_4 (z) = \psi^{\lambda-1} w_6 (\psi z) = \psi^{2(\lambda-1)} w_5 (\psi^2 z),
\end{gather}
\end{subequations}
where 
\beq
\psi \coloneqq e^{i\frac{2}{3} \pi}.
\eeq
One may also notice that, by considering 
\beq
\bar{w}_{\bar{j}} (z)\coloneqq(e^{-i \frac{\pi}{3}})^{\lambda-1} w_j (e^{-i \frac{\pi}{3}} z), 
\eeq
for $j=1,\ldots,6$, one may formally find basis sets also for the subluminal case. See figure 
\ref{fig:paths-ni}, right side.

In the following sections, we shall exploit the aforementioned mathematical formalism in order to study two models for 
the analogous Hawking effect in condensed matter system. We shall consider two subluminal cases, represented by the Corley 
subluminal model and the Hopfield model for dielectric media. In the companion paper \cite{belbecsw}, we shall deal with the superluminal 
case represented by Bose--Einstein condensates, and also the further subluminal case represented by surface waves.  

\section{Corley model: subluminal case}
\label{seccorley}

We refer mainly to Corley in the subluminal case, {\color{black} which is considered in \cite{corley,cpf,Coutant-thick}. 
It represents the simplest model one can consider in this field, and, differently from e.g. BEC and water waves, to be discussed in the 
companion paper~\cite{belbecsw}, it does not allow a variable speed of sound velocity $c(x)$. Furthermore, it cannot be related to the 
dielectric model which is discussed in the following section. As such, it is of limited physical interest, still we discuss it in our 
framework as a useful benchmark for our master equation and for our approximations. We shall not discuss the superluminal case 
for brevity. From} the action 
\beq
S=\frac{1}{2} \int d^2x [ ((\partial_t + v \partial_x)\phi)^2+\phi \frac{1}{k_0^2} \partial_x^4 \phi] 
\eeq
displayed in~\cite{corley-jacobson,corley}  one obtains by separation of variables the fourth order ordinary differential equation
\beq
\label{eqco}
\frac{1}{k_0^2}  \partial_x^4 \varphi + \biggl(1-\frac{v^2 (x)}{c^2}\biggr)  \partial_x^2 \varphi+ 2 \frac{v(x)}{c^2}  (i \omega -v'(x)) 
\partial_x \varphi- i \frac{\omega}{c^2} (i \omega -v' (x)) \varphi =0,
\eeq
where $v(x)$ is the velocity field and $v'(x)$ stands for its first derivative with respect to $x$, and we have restored the 
(constant) sound velocity $c$. 
In order to reproduce the features of the master equation above, one must consider the following 
choice of the scale parameter: we assume as a significant physical scale, as in~\cite{corley-jacobson,corley},
the scale $k_0$ associated with the nonlinearity.
By defining the dimensionless parameter\footnote{It might be questioned such a choice of expansion parameter, as other choices could appear as more natural, e.g. one could consider $\kappa$ (cf.~\eqref{kappa-def}) in place of $\omega$. Still, it can be verified that in the error estimates like e.g. in~\eqref{err-wkb} nothing substantial changes.}
\beq
\label{expa-p}
\epsilon \coloneqq \frac{\omega}{c k_0} ,
\eeq
and the dimensionless coordinate $\xi=x\omega/c$,~\eqref{eqco} becomes (with a small abuse of notation)
\beq
\label{eqcoad}
\epsilon^2  \partial_\xi^4 \varphi + \biggl(1-\frac{v^2 (\xi)}{c^2}\biggr)  \partial_\xi^2 \varphi+ 2 \frac{v(\xi)}{c}  \biggl(i -\frac {v'(\xi)}c\biggr) 
\partial_\xi \varphi+ \biggl(1 +i\frac{v' (\xi)}c\biggr) \varphi =0 .
\eeq
Assuming that $k_0\gg \omega/c$, we get $0<\epsilon^2 \ll 1$. Moreover, we have
\begin{subequations}
\begin{align}
p_3 (\xi,\epsilon) &= 1-\frac{v^2 (\xi)}{c^2} = p_{30}(\xi),\\
p_2 (\xi,\epsilon) &= 2 \frac{v(\xi)}{c} \biggl(i-\frac {v'(\xi)}c\biggr)  = p_{20}(\xi),\\
p_1 (\xi,\epsilon) &= 1 +i \frac {v' (\xi)}c = p_{10}(\xi).
\end{align}
\end{subequations}
There is no higher order contribution to the coefficients for this specific model (which is actually  
exceptional from this point of view). 
{\color{black} This is not true in the case of the other models we take into consideration herein and in the companion paper. We remark that the expansion parameter \eqref{expa-p} defining our limit of weak dispersion is the 
same as in \cite{cpf,Coutant-thick}}.

\subsection{The reduced equation} 
We notice that, in the limit $\epsilon\to 0$, one obtains the reduced equation, which we express in the original coordinates
\beq
\label{reduced-co}
\biggl(1-\frac{v^2 (x)}{c^2}\biggr)  \partial_x^2 \varphi+ 2 \frac{v(x)}{c^2}  (i \omega -v'(x)) 
\partial_x \varphi- i \frac{\omega}{c^2} (i \omega -v' (x)) \varphi =0 ,
\eeq
and, accordingly to~\cite{corley}, we assume $v(x)\leq 0$, so that the TP coincides with the solution of 
\beq
v(x) + c = 0 .
\eeq
In the neighbourhood of the TP we have
\beq
\label{linear}
v(x) \simeq -c+\kappa  x,
\eeq
where 
\beq
\label{kappa-def}
\kappa \coloneqq v' (x=0).
\eeq
The region where this approximation holds is called linear region henceforth. {\color{black} Notice that 
this is purposefully the same denomination as e.g. in \cite{cpf}.} 

The indicial equation for equation~\eqref{reduced-co} provides a vanishing root $\alpha_1=0$ and a 
non-vanishing one $\alpha_2= i \frac{\omega}{c \kappa}$, so that, being $\lambda=1-\alpha_2$, one gets 
\beq
\lambda = 1 - i \frac{\omega}{\kappa},
\eeq
which is not an integer number for any $\omega>0$.

\subsection{WKB approximation}
\label{corleywkb}

By now, we assume $x>0$, i.e. $|v|<c$, which means that the external region is taken into account. We put 
\beq
\varphi(\xi) = \exp \biggl( \frac{\theta(\xi)}{\epsilon} \biggr) \sum_{i=0}^\infty \epsilon^i y_i (\xi) ,
\eeq
and refer e.g.~to the presentation given in~\cite{holmes}.
To the lowest order, we obtain
\beq
\theta^{\prime 4} + \biggl(1-\frac{v^2}{c^2}\biggr) \theta^{\prime 2} =0, 
\eeq
whose solutions are $\theta'=0$ (multiplicity two), and 
\beq
\theta'_\pm = \pm i \sqrt{1-\frac{v^2}{c^2}}.
\eeq
Notice that, for $x<0$, being $|v|>c$, we obtain an exponentially increasing solution (called growing mode in \cite{cpf}), 
and a decaying solution.\\
We first take into account the latter solutions, and associate to them the so-called transport 
equation
\beq
\theta^{\prime 2} ( 6 \theta'' y_0+4 \theta' y'_0 +\theta^{\prime 2} y_1) + \biggl(1-\frac{v^2}{c^2}\biggr)  (\theta'' y_0+2 \theta' y'_0 +\theta^{\prime 2} y_1)
+ 2 \frac{v}{c} \biggl(i -\frac {v'}c\biggr) \theta' y_0 =0,
\eeq
and the next to leading order equation
\beq
\begin{split}
\label{eq:rec}
&\theta^{\prime 2} ( 6 \theta'' y_1+4 \theta' y'_1 +\theta^{\prime 2} y_2) + \biggl(1-\frac{v^2}{c^2}\biggr)  (\theta'' y_1+2 \theta' y'_1 +\theta^{\prime 2} y_2) + 2 \frac{v}{c} \biggl(i -\frac {v'}c\biggr) \theta' y_1 \\
&+5{\theta}^{\prime 2} y_0'' +\biggl(12 \theta'\theta'' +2\frac vc \biggl(i-\frac {v'}c\biggr)\biggr)y_0' +\biggl(3\theta^{\prime\prime2}+4\theta'\theta''+1+i\frac {v'}c\biggr)y_0 = 0.
\end{split}
\eeq
Going back to the original coordinates, we find the solutions
\begin{equation}
\begin{split}
\varphi_\pm (x) & = C \Biggl( \frac{1}{1-\frac{v^2 (x)}{c^2}} \Biggr)^{3/4} \exp ( \pm \frac{i}{\epsilon}\frac \omega{c} \int^x ds \sqrt{1-\frac{v^2 (s)}{c^2}} ) 
\exp (  i \frac{\omega}{c} \int^x ds \frac{v (s)}{c} \frac{1}{1-\frac{v^2 (s)}{c^2}}) \\
& \quad \times \Biggl( 1 + \epsilon C_1\pm \frac\epsilon2  \frac c\omega\int^x ds\ \frac 1{i\Bigl(1-\frac {v^2(s)}{c^2}\Bigr)^{\frac 32}}
\Biggl[ \frac 1{1-\frac {v^2(s)}{c^2}}\psi_1(s) +\psi_2(s)
\Biggr] +O(\epsilon^2)
\Biggr),
\end{split}
\end{equation}
where
\begin{align}
\psi_1(s)&=(i2\omega+3v'(s)) \frac {v^2(s)}{c^4} \left(\frac {15}4 v'(s)+\frac 34 i\omega\right)+\frac {v^2(s) v^{\prime2}(s)}{c^4}, \\
\psi_2(s)&=\frac {\omega^2}{c^2} -4i \frac {\omega v'(s)}{c^2} -\frac 72 \frac {v^{\prime2}(s)+v(s)v''(s)}{c^2}.
\end{align}
Omitting the terms of order $\epsilon$, the solutions correspond to the high wavenumber $k_\pm$ solutions appearing in~\cite{corley,cpf}.
$C$ is a normalization constant which, as in~\cite{corley}, we can put equal to one. 
$C_1$ is a second integration constant that also can be considered of order one. The $O(\epsilon)$ terms allow us to determine the conditions under which our approximations remains good. Since the integral diverges when $x\to 0$,
this approximation fails at the TP. Still, it is assumed to hold in the linear region. 
When $x\to\infty$, the integral part of the order $\epsilon$ terms goes like
\begin{equation}
\sim \mp i \frac 12  \frac 1{\Bigl(1-\frac {v_{r}^2}{c^2}\Bigr)^{\frac 32}} \Biggl(1- \frac 32 \frac {v_{r}^2}{c^2} \frac 1{1-\frac {v_{r}^2}{c^2}}\Biggr)  \frac {\omega^2}{c^2} \frac x{k_0},
\end{equation}
where $-c<v_{r}<0$ is the value assumed by $v$ far from the TP.
We observe that the validity of the approximation requires 
\begin{equation}
\label{err-wkb}
x\ll k_0 \frac {c^2}{\omega^2}\left(1-\frac {v_{r}^2}{c^2}\right)^{\frac 52} = \frac 1{\epsilon} \frac c\omega \left(1-\frac {v_{r}^2}{c^2}\right)^{\frac 52}.
\end{equation}
Since, as in~\cite{corley}, we are interested in very low frequencies, $\omega\sim0$, this is not a strong restriction at all, at least if the asymptotic velocity is not too close to $-c$.

It is also interesting to write the leading terms of $\varphi_\pm (x)$ in the linear region: 
\beq
\varphi_\pm (x) \simeq \biggl(\frac{2\kappa}{c} x\biggr)^{-3/4} x^{-\frac{i\omega}{2\kappa}}\exp \left( \pm \frac{i}{\epsilon} \frac{2}{3} 
\sqrt{\frac{2\kappa}{c}} x^{\frac{3}{2}}\right).
\eeq
Two further solutions occurring when $\theta'=0$  can be obtained  from the reduced equation.  The corresponding momenta are indicated, for a better comparison with~\cite{corley}, as $k_{\pm s}$ (in literature one finds also the following correspondence: $k_{+s} \mapsto k_u$, $k_{-s} \mapsto k_v$).
In order to maintain the same order of approximation in our WKB expansion, one would need exact solutions, in order to avoid the introduction of a further expansion parameter. Nevertheless, we can appeal to the general features of the equation itself.
Indeed, we obtain near the regular singular point $x=0$ (our TP) the following series expansions
\begin{align}
\varphi_{-s} (x) &= 1+\sum_{n=1}^{\infty} c_n x^n,\\
\varphi_{+s} (x) &= x^{i \frac{\omega}{\kappa}}  \biggl( 1+\sum_{n=1}^{\infty} d_n x^n\biggr) .
\end{align}
By comparing, as in~\cite{corley,cpf}, the behaviour of the above four solutions in the linear region where~\eqref{linear} holds, with the solutions near the TP (to be discussed in the following subsection), one finds both thermality and the grey body factor.

It is useful to provide approximate solutions of the reduced equation even for large $x$ (in the external region 
with respect to the black hole). It is easy to show that for large $x$ in the above sense we have $v(x)\sim$ const, and then $v'=0$.
As a consequence, e.g.~under the conditions of theorem 1.9.1 of~\cite{eastham}, we get as $x\to \infty$
\begin{align}
\varphi_{-s} (x) &\sim \exp ( -i {\omega} \frac{1}{c-v_{r}} x),\\
\varphi_{+s} (x) &\sim \exp ( i {\omega} \frac{1}{c+v_{r}} x), 
\end{align}
and this completes our asymptotic basis of solutions together with $\varphi_{-} (x)$ and $\varphi_{+} (x)$.
As useful interpolating formulas (WKB-like, but they cannot be rigorously obtained by using the 
$\epsilon$-expansion as in the above framework) we could also use 
\begin{align}
\varphi^{int}_{-s} (x) &\sim \exp ( -i {\omega} \int^x dy \frac{1}{c-v (y)} ),\\
\varphi^{int}_{+s} (x) &\sim \exp ( i {\omega} \int^x dy \frac{1}{c+v (y)}), 
\end{align}
which still display the correct behaviour both in the linear region and in the asymptotic one.

For $x<0$, the reduced equation provides us two further solutions 
\begin{align}
\varphi_{d} (x) &= 1+\sum_{n=1}^{\infty} e_n x^n,\\
\varphi_{l} (x) &= x^{i \frac{\omega}{\kappa}}  \biggl( 1+\sum_{n=1}^{\infty} f_n x^n\biggr) , 
\end{align}
with the asymptotic behaviour 
\begin{align}
\varphi_{d} (x) &\sim \exp ( -i {\omega} \frac{1}{c-v_{l}} x),\\
\varphi_{l} (x) &\sim \exp ( i {\omega} \frac{1}{c+v_{l}} x), 
\end{align}
with $\lim_{x\to -\infty} v(x) =: v_l <-c<0$. These solutions correspond to left-moving modes in the superluminal region, and they are the 
only propagating modes in that region. We notice that the mode $\varphi_{l} (x)$ is a negative-norm mode.\\
{\color{black} We remark that the modes we obtain are the same as in \cite{corley,cpf}, albeit obtained through a different approach 
to the WKB approximation. This confirms also  at this level the performance of our general framework.}

\subsection{Approximation near the turning point}
\label{corleynhe}

\begin{figure}[hbtp!]
\includegraphics[scale=0.55]{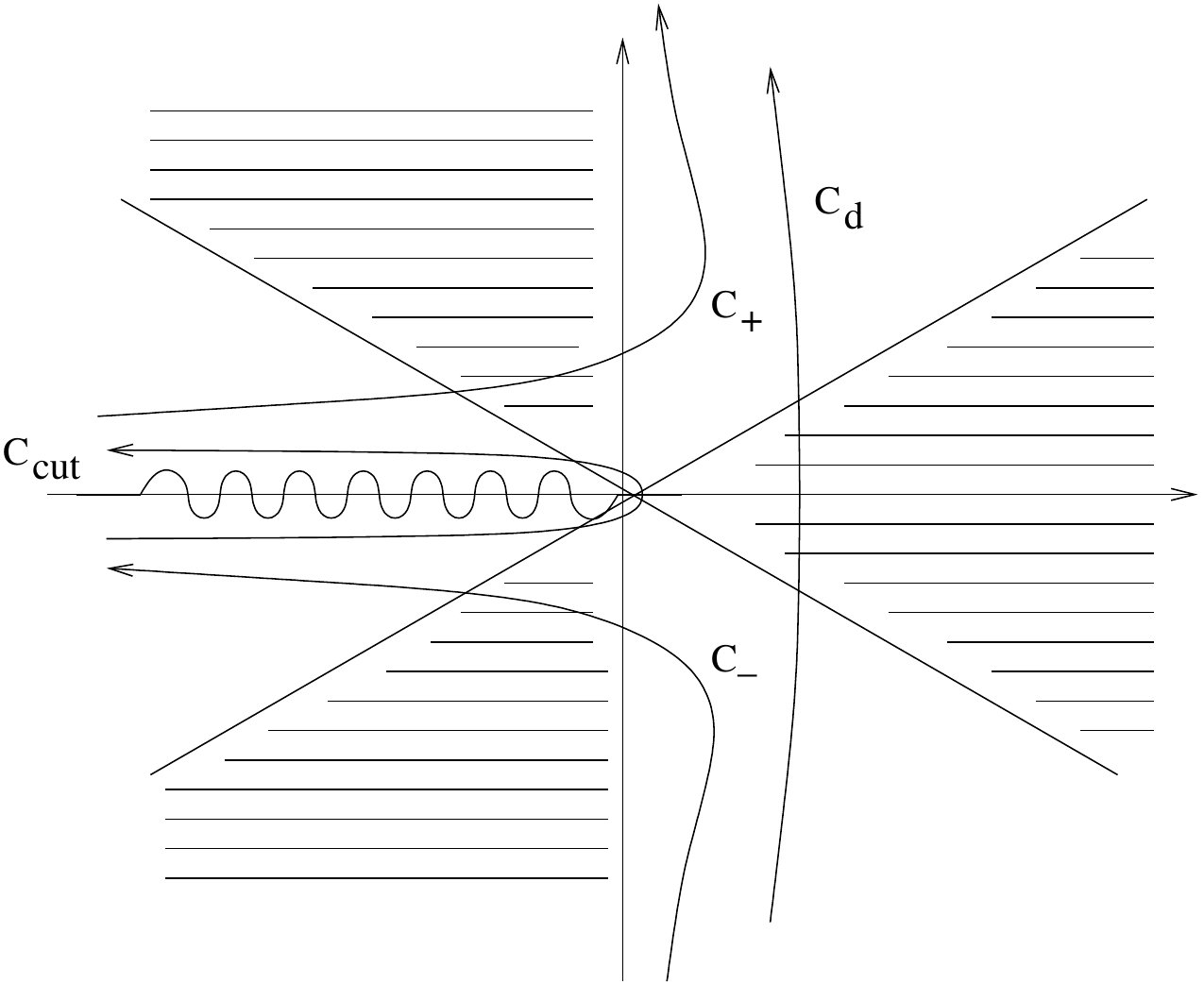} 
\caption{Paths used in the subluminal case in Corley's work~\cite{corley}. $C_\pm$ correspond to the dispersive modes, 
$C_{cut}$ to the Hawking mode, and $C_d$ to the decaying mode. The last mode is the one in the inner region $x<0$. 
As remarked by Corley~\cite{corley}, $C_d$ can be deformed in the paths $C_+$, $C_-$, $C_{cut}$.}
\label{fig:sub1}
\end{figure}

We first point out that, for the present case, we have\footnote{We use~\eqref{change-var} working with the coordinate $x$ in place of $\xi$.}
\beq
\label{zx-corley}
z= \biggl(\frac{2\kappa}{c}\biggr)^{1/3} \epsilon^{-2/3} x,
\eeq
and we choose to construct directly the relevant physical states by exploiting the method of the steepest descents~\cite{olver,wong,miller}.
The analysis proceeds as in the original paper by Corley ~\cite{corley},
with the relevant difference that a different parameter of the asymptotic expansion is proposed (e.g.~in~\cite{corley} the non-linearity scale $k_0$, which plays a fundamental role in our analysis, is put equal to one); 
furthermore, the near horizon equation allows to take into account the $-s$-mode, albeit in the form of a constant solution, which still matches the WKB behaviour in the matching region.
A different but rigorous tool for evaluating the branch cut contribution (see below) is exploited.
In previous literature, starting from \cite{corley}, the so-called boundary condition for the subluminal case required a decaying mode beyond the horizon ($x<0$), described by a path in the complex plane that can be deformed into the ones in 
the external region. The mathematical root of this condition will be discussed below. 
From a physical point of view, this condition fixes the relative amplitudes of the involved modes near the turning point. Strictly speaking, it is not a boundary condition 
in itself, but it indicates how the modes involved in the process at hand actually participate to the process itself. 
Figure~\ref{fig:sub1} amounts to the diagram introduced by Corley~\cite{corley}, in which the homotopic deformation of the decaying mode for $x<0$ gives the 
modes with momenta $k_+$, $k_-$, $k_{+s}$ appearing in the external region ($x>0$). They represent the high momentum 
incoming modes $k_\pm$, one of which having negative norm ($k_-$), and the outgoing Hawking mode $k_{+s}$. Since these modes 
must implement such a diagram near the TP, they participate with the same relative amplitude to the scattering process. 
The fourth mode, i.e. the short momentum regular mode,  $k_{-s}$, corresponds to a solution in the Laplace space 
(or, equivalently, in the Fourier space) that is of different nature. 
Therefore, it cannot be included in the diagram as a mode resulting from the homotopic deformation of the decaying mode. 
{\color{black} With respect to the analysis 
carried out in \cite{cpf,Coutant-thick}, which is instead based on a Fourier transform analysis, the diagrams involved in the Hawking effect 
are the same as in Figure~\ref{fig:sub1}.} 
In order to match the WKB solutions, we are interested in an asymptotic expansion for large $z$ (notice that this can be obtained also by leaving $x$ suitably small in order to allow the linear approximation hold true).

The $k_\pm$ contribution can be evaluated by means of the saddle point approximation, as well as the 
aforementioned decaying mode. We need the following formal expression
\beq
w_j(z) = \frac{1}{2\pi i} \int_{C_j} dt\ t^{\lambda -2} \exp (z t+ \frac{1}{3} t^3),
\eeq
which,  by putting $t=\sqrt{|z|} u$, can be rewritten as 
\beq
w_j(z) = \frac{1}{2\pi i} |z|^{\frac{\lambda-1}{2}} I_j (z),
\eeq
where 
\beq
I_j (z) = \int_{\bar{C}_j} du\ g(u) \exp (|z|^{3/2} h_\pm (u)),
\eeq
and
\begin{align}
g(u) & \coloneqq u^{\lambda-2},\\
h_\pm (u) & \coloneqq \pm u+\frac{u^3}{3} ;
\end{align}
here $\pm=\sign (x)$ and $C_j$ are the paths defined in~\cite{corley}. {\color{black} Cf. also \cite{cpf}.} 
We have for $x<0$ the decaying mode passing through the saddle point $u=1$ 
\beq
\label{decay-sub}
w_{decaying} (z)\simeq \frac{1}{2 \sqrt{\pi}} |z|^{-\frac{i\omega}{2\kappa}-\frac{3}{4}} e^{-\frac{2}{3} |z|^{3/2}}. 
\eeq
The other saddle point $u=-1$ corresponds to the growing mode (which diverges at infinity), whose coefficient in the scattering 
matrix is zero (cf. e.g. \cite{cpf,coutant-bdg}){\color{black}\footnote{{\color{black}The growing mode could still give some contributions in other 
context, see e.g. \cite{liberati}. We thank the anonymous referee for pointing out this.}} }

For $x>0$ we have the modes $k_\pm$ in correspondence of the steepest descents passing through the saddle points 
$u_\pm = \pm i$, and we get 
\begin{align}
\label{saddle-p-sub}
w_+ (z) \simeq & \frac{1}{2 \sqrt{\pi}} e^{-\frac{3}{4}\pi i} e^{\frac{\pi \omega}{2 \kappa}} |z|^{-\frac{i\omega}{2\kappa}-\frac{3}{4}} e^{i\frac{2}{3} |z|^{3/2}}, \\
w_- (z) \simeq & \frac{1}{2 \sqrt{\pi}} e^{\frac{1}{4}\pi i} e^{-\frac{\pi \omega}{2 \kappa}} |z|^{-\frac{i\omega}{2\kappa}-\frac{3}{4}} e^{-i\frac{2}{3} |z|^{3/2}}.
\label{saddle-m-sub}
\end{align}
It is nice to notice that, thanks to relation~\eqref{zx-corley}, the amplitude of the decaying mode and of the 
$k_\pm$ modes above are proportional to $\sqrt{\epsilon}$ and then vanish as $\epsilon\to 0$, as expected. 
We can also provide a bound on the error occurring in neglecting higher order contributions to the saddle point  
approximation. Following e.g.~\cite{olver} we find 
\beq
x^{3/2} \gg \frac{1}{k_0} \sqrt{\frac{c}{2 \kappa}} \frac{1}{8} \left( \frac{1681}{36}+4 \frac{\omega^4}{\kappa^4}+
\frac{110}{3}  \frac{\omega^2}{\kappa^2}\right)^{1/2}.
\eeq
As to the ratio $\omega/\kappa$, it is known that the Hawking effect is mostly peaked for $\omega \simeq \kappa$. As well known, there is also 
a maximal value of $\omega$ beyond which no Hawking effect occurs. See the following subsection for more details. 

As to the cut contribution, it represents the Hawking mode, as well known. It is remarkable that the branch cut  lies along a steepest descent.
Indeed, we have that for the subluminal case the imaginary part of $u+u^3/3$ is $b(1+a^2-b^2/3)$, where $u=a+ib$.
As a consequence, $b=0$ is a steepest descent line.
This allows us to compute the cut contribution along the lines suggested in~\cite{miller}, chapter 4, section 4.8, finding thus
\beq
\label{cut-sub}
w_{cut} (z) \simeq -\frac{1}{i \pi} |z|^{i \frac{\omega}{k}} \Gamma \Bigl(-i\frac{\omega}{\kappa}\Bigr) \sinh (\frac{\pi \omega}{\kappa}), 
\eeq
which coincides (apart for the factor $2\pi i$ we introduced) with the approximation given in~\cite{corley}, but on more rigorous grounds. 
{\color{black} This result is compatible with the analogous one obtained in \cite{cpf}, with the difference that the Fourier transfom formalism 
is adopted  and a dominated convergence  must be used therein}.

For $x<0$, it is easy to realize that the constant solution still appears.  
And one may also simply consider the 
contribution (\ref{cut-sub}) by choosing a suitable analytical continuation for $x<0$. It turns out that, by choosing the branch 
where $-1=e^{-i\pi}$, the further solution one obtains 
\beq
\label{cut-sub1}
w_{cut-l} (z) : \simeq -\frac{1}{i \pi}  e^{\pi \frac{\omega}{\kappa}} z^{i \frac{\omega}{k}} \Gamma \Bigl(-i\frac{\omega}{\kappa}\Bigr) \sinh (\frac{\pi \omega}{\kappa}), 
\eeq
is such that it corresponds to the Hawking partner, living on a different branch (cf. also \cite{corley}); furthermore, one is enabled to obtain the so-called mode which straddles the horizon \cite{visser-ess}. See the following subsection. 

\subsection{Matching: complete solutions}
A careful comparison with the WKB expansion displayed in the previous section provides us the connection formulas (cf.~the so-called central connections 
in~\cite{ni}).  It has to be remarked that, as a consequence of the Corley's black hole boundary condition, in the external region near the turning point we have
\beq
\label{nearhorizon-1}
\phi (x,t) = \phi_1 (x,t) + \phi_2 (x,t) + \phi_3 (x,t)+h \phi_4 (x,t), 
\eeq
with $\phi_1 \mapsto w_+$, $\phi_2 \mapsto w_-$, $\phi_3 \mapsto w_{cut}$, $\phi_4 \mapsto 1$ and where 
$h$ remains undetermined by adopting the diagram of figure~\ref{fig:sub1}.\\
Therefore, the modes corresponding to $w_\pm,w_{cut}$ enter with the same amplitude in the scattering matrix. Instead, for the fourth constant mode, room is left for a different amplitude, as indicated by the 
factor $h$ in front of it. Eventually, $h$ might even be set equal to zero, see also the discussion below equation \eqref{connection}. From a mathematical point of view, the solutions $w_\pm,w_{cut},w_{decay}$, 
for $z=0$, where they are regular, as a consequence of Cauchy's  Theorem, satisfy 
\beq
\label{hor-cond}
w_+(0) + w_-(0) +w_{cut}(0) =w_{decay}(0).
\eeq
For what concerns strictly the problem of fixing the relative amplitudes of the respective modes, this amounts to the above boundary condition stated by considering  
modes on different sides of the real turning point. Condition \eqref{hor-cond} works as well as the original condition by Corley. 

A complete description of the matching  is described in Appendix \ref{matching}. By comparing with the WKB solutions 
again in the matching region we find
\begin{equation}
\begin{split}
\label{connection}
\phi (x,t) & = 
e^{-\frac{3}{4}\pi i}  \frac{e^{\frac{\pi \omega}{2\kappa} }}{2 \sqrt{\pi}} \Bigl(\frac{2 \kappa}{c}\Bigr)^{-\frac{i\omega}{6  \kappa}
+\frac{1}{2}}
\epsilon^{\frac{i\omega}{3  \kappa}+\frac{1}{2}}\varphi_+ (x,t)\\
& \quad + e^{\frac{1}{4}\pi i}  \frac{e^{-\frac{\pi \omega}{2\kappa} }}{2 \sqrt{\pi}} \Bigl(\frac{2 \kappa}{c}\Bigr)^{-\frac{i\omega}{6  \kappa}+\frac{1}{2}}
\epsilon^{\frac{i\omega}{3  \kappa}+\frac{1}{2}} \varphi_- (x,t) \\
& \quad - \frac{\sinh (\frac{\pi \omega}{\kappa}) }{\pi i} \Gamma \Bigl(-  \frac{i\omega}{\kappa}\Bigr)
\Bigl(\frac{2 \kappa}{c}\Bigr)^{\frac{i\omega}{3  \kappa}} 
\epsilon^{-\frac{2i\omega}{3  \kappa}}
\varphi_{+s} (x,t)+h \varphi_{-s} (x,t). 
\end{split}
\end{equation} 
$h$ is still undetermined. The fact that the fourth mode $\varphi_{-s} (x,t)$ is not involved in the 
Corley's black hole boundary condition, suggests the following interpretation: it does not participate the process of Hawking particle production very near the 
TP, but it still might participate at a subsequent stage when scattering on the geometry depletes the flux of Hawking particles by `barrier reflection'.  
This is what consistently appears to hold true for the model at hand, as also a direct calculation of the 
emitted flux confirms.\\
In literature, there exist two models where the fourth mode appear in the Corley diagram, see \cite{linder} and \cite{hopfield-kerr}, where the fourth mode solution near the horizon has the same functional dependence of the other three solutions, 
and $h=1$ occurs. Homotopic deformation from the decaying mode involves also the fourth mode, and its direct contribution to the grey-body factor appears \cite{linder}. 
On the grounds of the comparison with these models, being the fourth mode not present in the Corley's 
diagram, {\color{black} the absence of the mode leaves $h$ undetermined; we can also infer that there is no contribution to $h$ for what strictly concerns the pair-creation process at least at the leading order. We shall discuss the problem further in the following}.  

As to the modes $d,l$ in the black hole region, their matching is analogous to the one described above. 
The mode $d$, as discussed above, may be considered, together with its 
counterpart $-s$ on the external side of the horizon, a single mode representing the particle entering the hole, and, as such, it passes without any relevant effect. Of 
course, it can also participate to the whole scattering process for Hawking particles as the backward mode originated from scattering on the geometry 
of Hawking particles.   
The other mode $l$ can be again straightforwardly matched with its WKB part, and together with the $+s$ mode one may 
define the so-called straddle mode:
\beq
\phi_{straddle} (x,t) \coloneqq \phi_{+s} (x,t) \theta (x) +\phi_{l} (x,t) \theta (-x),
\eeq
where $\theta(x)$ is the Heaviside function.  This mode, starting from the matching regions on both sides of the 
turning point, is composed by the Hawking mode on the external side, and of the Hawking partner on the black hole side. 
It contains a Planckian distribution of Hawking modes in the external region \cite{visser-ess}. With respect to the standard case, 
there is of course a near-horizon regular part of the mode which is missing in the standard black hole case. {\color{black} See 
also the discussion in \cite{Coutant-thick}.}

 It may be noticed that, due to the transformation defined in~\eqref{yw}, each solution in the near horizon approximation should be 
multiplied by an overall factor $\epsilon^{4/3}$. We can reabsorb this factor in the normalization. We shall adopt this convention henceforth in all the 
models we take into consideration. 

\subsection{Thermality}

As usual, for thermality one may verify that 
\beq
\frac{|J_x^{-}|}{|J_x^{+}|} = e^{-\beta \omega},
\eeq
where 
\beq
\beta\coloneqq \frac{2 \pi}{\kappa}
\eeq 
is the inverse Hawking temperature. 
We stress that, in this sense, thermality is unaffected by the still undetermined value of $h$. The current conservation provides
\beq
|J_x^{+s}| = |J_x^{+}|-|J_x^{-}|+ |J_x^{-s}|, 
\label{currents}
\eeq
which amounts to the usual relation between the Bogoliubov coefficients 
involved in the process. If we separate each contribution by $|J_x^{+s}|$ we obtain the square modulus of the 
amplitudes in \eqref{amplitudes}. 

We note that there is also the contribution of 
the regular mode $-s$, which is missing in the near-horizon diagram~\ref{fig:sub1}. 
The subtle point is that 
{\it a priori}, the flux at infinity of the Hawking mode $+s$ can be depleted because of 
scattering (reflection) on a potential barrier emerging as an effect of the geometry. 
It has 
nothing to do with the horizon itself, as in the well-known astrophysical case: in four 
dimensions, e.g.~a scalar particle on the Schwarzschild background is affected by the presence of a 
centrifugal barrier in the external region of the black hole (apart for $l=0$ modes), which can 
reflect back to the horizon the Hawking quanta. Of course, in 2D Schwarzschild case this phenomenon 
is absent (no centrifugal contribution). 
{\color{black} This discussion is in agreement with the one carried out in \cite{cpf,Coutant-thick},  
where only scattering effects are present in the depletion of the 
Hawking flux.} 

\subsection{Grey-body factor} In order to get also the grey-body factor one must evaluate the ratio
\beq
\label{ratio-short}
R \coloneqq  \frac{|J_x^{-s}|}{|J_x^{+s}|},
\eeq
which indicates the fraction of particles reflected back, and then obtain 
the grey-body factor as  
\beq
\label{greybody}
\Gamma =1-R= 1 - \frac{|J_x^{-s}|}{|J_x^{+s}|} .
\eeq
In line of principle, one might deduce the grey-body factor from the direct calculation of 
\beq
|\beta_\omega|^2 \coloneqq \frac{|J_x^{-}|}{|J_x^{+s}|}=|\bar{C}_{-}|^2,
\eeq
which represents the number of created particles, as known (for the second equality cf. \eqref{cimeno}). In the case of the present model, one would obtain a perfectly Planckian 
spectrum with $\Gamma=1$, which implies $h=0$. Still, even if this route is viable, there is the risk of a poor approximation 
(as in the standard Hawking effect calculations).\\ 
We notice that fluxes in \eqref{greybody} are both calculated at $x=\infty$, which is the only asymptotic region available to both the modes at hand. 
{\color{black} Our strategy in the present framework for the calculation of the grey-body factor consists in taking account of the 
scattering contribution to the geometry 
simply by} studying  the reduced 
equation for the $\pm s$ modes, reducing it in the form of a Schr\"odinger equation. This might be obtained 
by means of a suitable variable transformation on the geometry associated with the reduced equation, 
which is the geometry of the analogue black hole, allowing to switch to Schwarzschild-like coordinates where 
the metric is diagonal and only a second order term in spatial derivatives appears. 
Indeed, the reduced equation, which 
valid in the WKB approximation, couples the short wavenumber modes $\pm s$ each other. Given a $+s$-mode entering from the 
part of the linear region, where the WKB approximation is valid, we are enabled to calculate 
\beq
R=:R_{reduced} \coloneqq  \left(\frac{|J_x^{-s}|}{|J_x^{+s}|}\right)_{reduced},
\eeq
with the fluxes computed asymptotically, using e.g.\footnote{Notice that this is not mandatory, cf. e.g. \cite{anderson-exact} for the BEC case.}  tortoise-like coordinate $\rho$ (see \eqref{tortoise} below), and with $|J_x^{-s}|$ measured 
at $\rho=-\infty$ (i.e. near the horizon, but still in a region where the WKB works well). That value would give a mechanism of interplay between the two short wavenumber modes, which should be taken properly into account. {\color{black} Cf. again the discussion in \cite{cpf,Coutant-thick}. 
Notice that, in general, the reduced equation has a quite involved form, 
and it is not easy to solve exactly, except for particular cases. As in the astrophysical case, it allows also further 
approximations with respect to the weak dispersion scheme we propose herein. Indeed, even a limit of low frequency 
can be adopted, as in the astrophysical case, without making difficult to ascertain if thermality is present, as thermality 
is anyway granted by the calculations above.}  {\color{black} There is also a further possible interpretation, indeed one may also choose to measure the flux of particles entering the horizon by measuring the flux of modes $d$ at $x=-\infty$, as the flux of entering particles generated by the back-scattering and measured by the static observer must 
coincide with the one of modes $d$ arriving at $x=-\infty$, so that $R={|J_x^{\ d}|}/{|J_x^{+s}|}$.}

To be more explicit,~\eqref{reduced-co} is of course equivalent to the Klein--Gordon equation 
\beq
\Box \phi (x,t) =0,
\eeq
on the curved background metric 
\beq
ds^2 = c^2 dt^2 - (v(x)\; dt - dx)^2,
\eeq
when static solutions $\phi (x,t) =e^{-i \omega t} \varphi (x)$ are considered. A standard coordinate 
transformation 
\beq
dt=d\tau-\frac{g_{01(x)}}{g_{00}(x)} dx,
\eeq
carries the metric in the diagonal Schwarzschild-like form 
\beq
ds^2 = \biggl(1-\frac{v(x)^2}{c^2}\biggr) c^2 {d\tau}^2 -\frac{1}{1-\frac{v(x)^2}{c^2}}{dx}^2, 
\eeq
so that, by choosing the tortoise-like coordinate
\beq
\label{tortoise}
\rho \coloneqq \int \frac{dx}{1-\frac{v(x)^2}{c^2}},
\eeq
one obtains the following Schr\"odinger-like equation 
\beq
\frac{1}{1-\frac{v(x(\rho) )^2}{c^2}} \biggl( \frac{d^2 \varphi (\rho)}{d\rho^2}+\omega^2 \varphi (\rho) \biggr) =0, 
\eeq
which amounts to a free equation in the external region. Therefore, there is no barrier, i.e.~no reflection, and the 
the grey body factor is trivially 
\beq
\Gamma = 1.
\eeq
As a consequence, $h=0$ and then, in this framework the model at hand is purely thermal, at least at the leading order in $\epsilon$.\\
As well known since 
former studies on the dispersive models, there exists a maximal frequency $\omega_{max}$ such that, for $\omega > \omega_{max}$, 
only two modes participates to the scattering process and the Hawking effect is no more present~\cite{corley-jacobson}. 
It is also known that $\omega_{max}$ is proportional to the dispersive scale $k_0$ both in the subluminal and in the superluminal cases~\cite{macher-wh,macher-bec}, and then it  goes to infinity in the limit as $k_0\to \infty$ (i.e.~as $\epsilon\to 0$). One has to expect that the spectrum is truncated at $\omega_{max}$ for non-zero values of $\epsilon$. 

\section{The dielectric case}

This case is just more tricky, since one has to deal with a system of differential equations 
instead of a single equation. Indeed, in the so-called $\phi$--$\psi$ model~\cite{hawbook}, one has 
\begin{align}
{\mathcal L}_{\varphi\psi} = \frac{1}{2} (\partial_\mu \phi)(\partial^\mu \phi)+\frac{1}{2\chi \omega_0^2} \left[ (v^\alpha \partial_\alpha \psi)^2 - \omega_0^2 \psi^2 \right] + \frac{g}{c} (v^\alpha \partial_\alpha \psi) \phi,
\label{Lagrangian}
\end{align}
where $\phi,\psi$ play the role of electromagnetic field and polarization field respectively, $\chi$ plays the role of the dielectric susceptibility, $v^\mu$ is the usual four-velocity vector of the dielectric, $\omega_0$ is the proper frequency  of the medium, and $g$ is the coupling constant between the fields. We get 
the system
\begin{align}
\label{phi-eq}
\Box \phi - \frac{g}{c} (v^\mu \partial_\mu \psi) & =0, \\
\label{psi-eq}
\left(\frac{1}{\chi \omega_0^2} (v^\mu \partial_\mu)^2 +\frac{1}{\chi}\right) \psi+  
\frac{g}{c} (v^\mu \partial_\mu \phi) & =0 . 
\end{align}
For simplicity, we put $g=1$ in what follows (as this parameter, introduced in~\cite{hopfield-hawking}, is no more 
necessary herein). 
We proceed as in~\cite{hopfield-kerr}, by considering that the spatial dependence appears in $\chi$ and in $\omega_0$ 
in such a way that $\chi \omega_0^2=$ const. Cf. also~\cite{hawbook}, chapter 10.  
In this case, we identify 
\beq
\epsilon^2 \coloneqq \frac{1}{\chi \omega_0^2} ,
\eeq
as the small parameter occurring in the problem.

\subsection{A separated equation for $\psi$}

We apply the operator $\Box$ on the left of equation~\eqref{psi-eq} (cf.~\cite{linder}),
and by taking into account the stationary case, where $\phi = \varphi (x) e^{i \omega t}$, 
$\psi = f(x) e^{i \omega t}$ are in the kernel of the operator 
$[\Box,v^\mu \partial_\mu]$, one obtains the following fourth order ordinary differential 
equation:
\begin{equation}
\begin{split}
&-\epsilon^2 \partial_x^4 f - 2 i \epsilon^2 \frac{\omega}{v} \partial_x^3 f+ \frac{1}{\chi \gamma^2 v^2}
\left( -\biggl(1- \chi \gamma^2 \frac{v^2}{c^2}\biggr) +\epsilon^2 \chi \omega^2\right) \partial_x^2 f+
2 \biggl(i \frac{\omega}{v} \frac{1}{c^2} (1 - \epsilon^2 \omega^2) - \frac{1}{\gamma^2 v^2} \biggl(\partial_x \frac{1}{\chi}\biggr) \biggr) \partial_x f \\
& + \left( \epsilon^2 \frac{\omega^4}{v^2 c^2}-   \frac{1}{\gamma^2 v^2} \biggl(\partial_x^2 \frac{1}{\chi}\biggr)
- \frac{\omega^2}{\chi\gamma^2 v^2 c^2} - \frac{\omega^2}{c^2 v^2} \right) f =0.
\end{split}
\end{equation}
In order to obtain a form reproducing the original master equation~\eqref{orso}, we need a further step:
we define $f(x) = h(x) \zeta(x)$, with
\beq
h(x) = A \exp (-i \frac{\omega}{2 v} x),
\eeq
where $A$ is a constant. $h(x)$ is chosen such that the third order term vanishes, and the procedure is analogous to 
the Liouville transformation which eliminates the first order term in a second order linear ordinary differential equation. 
This leads to the following quartic equation, which is just of the type `Orr--Sommerfeld' in the sense 
described in the previous sections
\beq
\label{orso-psi}
\begin{split}
&-\epsilon^2 \partial_x^4 \zeta + \biggl[ -\frac{1}{\chi \gamma^2 v^2} \biggl(1- \chi \gamma^2 \frac{v^2}{c^2}\biggr) 
+\epsilon^2 \frac{1}{\gamma^2 v^2}  \biggl(1-\frac{3}{2} \gamma^2\biggr) \omega^2\biggr] \partial_x^2 \zeta   \\
& +\biggl(  i \frac{\omega}{v} 
\frac{1}{\chi \gamma^2 v^2} 
\biggl( 1+\chi \gamma^2 \frac{v^2}{c^2} \biggr) -2 \frac{1}{\gamma^2 v^2} \biggl(\partial_x \frac{1}{\chi}\biggr) -i \epsilon^2 \frac{\omega^3}{v c^2}\biggr) \partial_x \zeta \\
& +
\biggl[\frac{1}{ \gamma^2 v^2} \biggl( i \frac{\omega}{v} \biggl(\partial_x \frac{1}{\chi}\biggr) -\biggr(\partial_x^2 \frac{1}{\chi}\biggr)\biggr) 
+ \frac{1}{ \gamma^2 v^2} \biggl( \frac{1}{4 \chi} \frac{\omega^2}{v^2} \biggl(1-\chi \gamma^2 \frac{v^2}{c^2} \biggr) 
-\frac{\omega^2}{\chi c^2} \biggl) \\
&
+\epsilon^2 \biggl(\frac{\omega^4}{v^4} \biggl(-\frac{1}{16}+\frac{1 v^2}{4 c^2}\biggr)\biggr) \biggr] \zeta =0.
\end{split}
\eeq
The TPs occur for 
\beq
1-\chi(x) \gamma^2 \frac{v^2}{c^2} =0,
\eeq
and we consider only the black hole solution.

The reduced equation is 
\beq
\frac{1}{\chi \gamma^2 v^2}  \biggl(1- \chi \gamma^2 \frac{v^2}{c^2} \biggr)
\partial_x^2 \zeta
-\frac{1}{ \gamma^2 v^2} \biggl(  i \frac{\omega}{v} 
\frac{1}{\chi } 
\biggl( 1+\chi \gamma^2 \frac{v^2}{c^2} \biggr) -2\biggl(\partial_x \frac{1}{\chi}\biggr) \biggr) \partial_x \zeta +
[\cdots] \zeta =0,
\eeq
where the limit $\epsilon\to 0$ is taken and $[\cdots]$ is a contribution readable just from \eqref{orso-psi}, and which does not 
participate to the indicial equation, whose roots are
\beq
\alpha_1 =0, \qquad \alpha_2 = -1 -i \frac{\omega c}{\gamma^2 v^2 n'}.
\eeq
$n$ is the refractive index, which is defined such that 
\beq
n^2 = 1+ \chi ;
\eeq
then we find
\beq
\lambda=2 +i \frac{\omega c}{\gamma^2 v^2 n'}. 
\eeq
Thanks to such a knowledge, one is able to find out the behaviour of $\psi$ in all regions of interest, and 
in particular in the matching region.

\subsubsection{WKB approximation}

As in subsection~\ref{corleywkb}, we put
\beq
\zeta (x) = \exp ( \frac{\theta(x)}{\epsilon} ) \sum_{i=0}^\infty \epsilon^i y_i (x) ,
\eeq
and obtain
\beq
\theta^{' 4} + \frac{1}{\chi \gamma^2 v^2}\biggl(1- \chi \gamma^2 \frac{v^2}{c^2}\biggr) \theta^{' 2} =0, 
\eeq
whose solutions are $\theta'=0$ (multiplicity two), and for $x>0$
\beq
\theta'_\pm = \pm i \frac{1}{\sqrt{\chi} \gamma v}\sqrt{1-\chi \gamma^2 \frac{v^2}{c^2} }.
\eeq
The latter solutions are associated with the transport equation
\beq
y'_0+\frac{1}{(1-\chi \gamma^2 \frac{v^2}{c^2})} \biggl[ -\frac{1}{4} \frac{\chi'}{\chi} +
i \frac{\omega}{2 v} \biggl(1+ \chi \gamma^2 \frac{v^2}{c^2}\biggr) \biggr] y_0=0,
\eeq
and the next to leading order equation
\beq
\begin{split}
y'_1+\frac{1}{1-a} \biggl[ -\frac{1}{4} \frac{\chi'}{\chi} +i \frac{\omega}{2 v} (1+ a) \biggr] y_1=&\mp i\frac {\sqrt \chi\ v\gamma}{(1-a)^{\frac 32}}\biggl[ \frac 94 \frac {\chi''}\chi -\frac 1{1-a} \frac {37}{16} \frac {\chi^{\prime2}}{\chi^2}
+i \frac \omega{v} \frac {\chi'}\chi \frac {4+7a}{1-a} \\
& +\frac {\omega^2}{v^2} \frac a{1-a}  +\frac {\omega^2}{c^2} -\frac {\omega^2}{c^2} \biggl(\frac 32 \gamma^2-1\biggr) \frac {\chi}a (1-a) \biggr] y_0,
\end{split}
\eeq
where we have defined
\beq
a(x) \coloneqq \chi (x) \gamma^2 \frac{v^2}{c^2}.
\eeq
Solutions are of the form 
\begin{align}
y_0 (x) & = B \chi (x)^{1/4} (1-a(s))^{-1/4} e^{-i \frac{\omega}{2 v} \int^x ds \frac{1+a(s)}{1-a(s)}}, \\
\begin{split}
y_1(x) & = y_0(x) \biggl\{ D \mp i\int^x ds\ \frac {\sqrt \chi\ v\gamma}{(1-a(s))^{\frac 32}}\biggl[ \frac 94 \frac {\chi''(s)}{\chi(s)} -\frac 1{1-a(s)} \frac {37}{16} \frac {\chi^{\prime2}(s)}{\chi^2(s)}
+i \frac \omega{v} \frac {\chi'(s)}{\chi(s)} \frac {4+7a(s)}{1-a(s)} \\
& \quad +\frac {\omega^2}{v^2} \frac {a(s)}{1-a(s)}  +\frac {\omega^2}{c^2} -\frac {\omega^2}{c^2} \biggl(\frac 32 \gamma^2-1\biggr) \frac {\chi(s)}{a(s)} (1-a(s)) \biggr] \biggr\},
\end{split}
\end{align}
where $B$ and $D$ are constants.
Then we obtain the high momentum modes 
\beq
\zeta_\pm (x) = e^{\pm \frac{i}{\epsilon} \frac{1}{\gamma v} \int^x ds \frac{1}{\sqrt{\chi (s)}} 
\sqrt{1-a(s) }} \bigl(y_0 (x)+\epsilon y_1(x) +O(\epsilon^2)\bigr).
\eeq
If we require $\epsilon y_1<y_0$ in the region where $\chi$ is essentially constant, we get the restriction
\begin{equation}
x<\frac {\omega_0 v}{\omega^2} (1-a_{as})^{\frac 52},
\end{equation}
where $a_{as}$ is the asymptotic value of $a(x)$. Since typically $1-a_{as}\sim 10^{-2}$, by $v/\omega=\lambda/(2\pi)$ we can also write
\begin{equation}
 x< \frac {\omega_0}\omega \frac \lambda{2\pi}  10^{-5}.
\end{equation}
This implies that the approximation is valid for frequencies such that $\omega\ll \omega_0$.

Near the TP one obtains 
\beq
|f_\pm (x)| \propto x^{-1/4},
\eeq
as found in~\cite{hopfield-kerr} for the electromagnetic case and in the $\phi$--$\psi$ model, see~\cite{hawbook}, chapter 10.
 For $x<0$, the solutions with $\theta'\not =0$ are exponentially decaying (decaying mode) and growing (growing mode) 
respectively.

Two further solutions occur from the reduced equation, when $\theta'=0$.
We find near the regular singular point 
$x=0$ (our TP) the following series expansions for $x>0$: 
\begin{align}
\zeta_{-s} (x) &= 1+\sum_{n=1}^{\infty} c_n x^n,\\
\zeta_{+s} (x) &= x^{-1-i \frac{\omega c}{\gamma^2 v^2 n'}}  \biggl( 1+\sum_{n=1}^{\infty} d_n x^n\biggl) .
\end{align}
Still, we get the same behaviour near the TP as calculated in~\cite{hopfield-kerr} for the electromagnetic case and in~\cite{hawbook}, chapter 10, for the simpler $\phi$--$\psi$ model
\begin{align}
|f_{-s} (x)| &\propto \text{const},\\
|f_{+s} (x)| &\propto x^{-1}.
\end{align}
As in the Corley model discussed in the previous section, we can also obtain for $x<0$  two further modes $d,l$ which 
asymptotically propagate towards $x=-\infty$. We don't provide details, as they are straightforward. 

\subsubsection{Approximation near the turning point}

Solutions near the TP have the following behaviour in the matching region, and we recall that $\lambda=2 +i \frac{\omega c}{\gamma^2 v^2 n'}:=2 -i \frac{\omega c}{\kappa}$, where
\beq
\kappa \coloneqq \gamma^2 v^2 |n'|
\eeq
amounts to the surface gravity of the dielectric black hole (see e.g.~\cite{belgiorno-prd}).
Being 
\beq
z=\biggl(\frac{2 \kappa}{v c^3}\biggl)^{1/3} \epsilon^{-2/3} x,
\eeq 
we can exploit the solutions we found in the previous section, as formally we have the same equation and 
then the same solutions (with different explicit values of $p'_ {30}(0)$ and of $\epsilon$). 
As a consequence, we obtain 
for $x<0$ the decaying mode in an analogous way as for~\eqref{decay-sub}, and it provides us the black hole boundary condition 
for the present model (which is subluminal, too). For $x>0$ we have the modes $k_\pm$ in correspondence of the steepest descents passing through the saddle points $u_\pm = \pm i$, i.e.  
\begin{align}
\label{saddle-p-psi}
w_+ (z) &\simeq \frac{1}{2 \sqrt{\pi}} e^{-\frac{3}{4}\pi i} e^{\frac{\pi \omega c}{2 \kappa}} |z|^{-\frac{i\omega c}{2\kappa}-\frac{3}{4}} e^{i\frac{2}{3} |z|^{3/2}}, \\
\label{saddle-m-psi}
w_- (z) &\simeq \frac{1}{2 \sqrt{\pi}} e^{\frac{1}{4}\pi i} e^{-\frac{\pi \omega c}{2 \kappa}} |z|^{-\frac{i\omega c}{2\kappa}-\frac{3}{4}} e^{-i\frac{2}{3} |z|^{3/2}}.
\end{align}
As to the cut contribution, we find 
\beq
\label{cut-die}
|w_{cut} (z)| \simeq \biggl| \frac{1}{i \pi} \Gamma \Bigl(1-i\frac{\omega c}{\kappa}\Bigr) \sinh (\frac{\pi \omega c}{\kappa}) \biggl|.
\eeq
It is easy to show that a matching is possible in the linear region, and thermality can be easily verified. Still, as the polarization field is substantially an `ancillary field' in the model, the really propagating field being the electromagnetic one, 
we prefer to calculate the matching and thermality of the spectrum by following a different route.

\subsection{A separated equation for $\phi$}

One might get an equation for $\phi$ as in~\cite{linder}, with the drawback of a tricky complication 
for dealing the limit as $\omega \to 0$.
Hence we prefer to proceed in a different way, and obtain 
a fourth order equation for $\phi$ from the original system of differential equations~\eqref{phi-eq} and~\eqref{psi-eq}.

Our trick is again to separate the variables in the comoving frame, with $\phi = \varphi (x) e^{i \omega t}$, 
$\psi = f(x) e^{i \omega t}$. A quartic equation is obtained as follows: we apply the operator 
$(i \omega +v \partial_x)$ to both the members of~\eqref{phi-eq}
\beq
\label{trick}
(i \omega +v \partial_x) \biggl( -\frac{\omega^2}{c^2} -\partial_x^2\biggr) \varphi = \frac{1}{c} \gamma (i \omega +v \partial_x)^2 f ;
\eeq
from~\eqref{psi-eq} one can isolate the term $f/\chi$ on the left side, and by finding $(i \omega +v \partial_x)^2 f$ from~\eqref{trick}
one obtains
\beq
f = -\frac{1}{c} \chi \gamma (i \omega +v \partial_x) \varphi -\chi \epsilon^2 \gamma c (i \omega +v \partial_x) 
\biggl( -\frac{\omega^2}{c^2} -\partial_x^2\biggr) \varphi.
\eeq
Then one can exploit equation~\eqref{phi-eq} on the separated variables 
\beq
\biggl( -\frac{\omega^2}{c^2} -\partial_x^2\biggr) \varphi = \frac{1}{c} \gamma (i \omega +v \partial_x) f ,
\eeq 
together with the above expression for $f$, and get the fourth order equation
\beq
\begin{split}
&-\epsilon^2 \gamma^2 v^2 \chi \partial_x^4 \varphi - \epsilon^2 \gamma^2 ( 2 i \omega \chi v +v^2 (\partial_x \chi)) \partial_x^3 \varphi\\
&-\biggl(1-\chi \gamma^2 \frac{v^2}{c^2} +\epsilon^2 \biggl( i \frac{\omega}{v} \gamma^2 v^2 (\partial_x \chi) -
\chi \omega^2 \biggr)\biggr)\partial_x^2 \varphi\\
&+\biggl[ \frac{1}{c^2} \chi \gamma^2 v^2 \biggl(2 i \frac{\omega}{v}+\frac{1}{\chi} (\partial_x \chi) \biggr) 
- \epsilon^2 \gamma^2 \biggl(2 i \frac{\omega^3}{c^2} \chi v + v^2 (\partial_x \chi) \frac{\omega^2}{c^2}\biggr)
 \biggr] \partial_x \varphi\\
&+\biggl[-\frac{\omega^2}{c^2} - \frac{1}{c^2} \chi \gamma^2 \omega^2+i v\frac{1}{c^2} \gamma^2 \omega (\partial_x \chi) +
\epsilon^2 \gamma^2 \chi \frac{\omega^4}{c^2}-\epsilon^2 i \gamma^2 v \frac{\omega^3}{v^2} (\partial_x \chi) \biggr] \varphi =0 .
\end{split}
\eeq
In order to eliminate the third order term, we put $\varphi= h(x) \eta(x)$,
and in this case the function $h(x)$ must satisfy the differential equation
\beq
4 h'+\biggl( 2 i \frac{\omega}{v}+\frac{1}{\chi} (\partial_x \chi) \biggr) h =0,
\eeq
whose solution is 
\beq
h = A \, \chi^{-1/4} e^{-i \frac{\omega}{2 v} x}. 
\eeq
Then one obtains the fourth order differential equation for $\eta$ in the desired form: 
\beq
\begin{split}
&-\epsilon^2 \gamma^2 v^2 \chi \partial_x^4 \eta -\biggl(1- \chi \gamma^2 \frac{v^2}{c^2}+O(\epsilon^2)\biggr) \partial_x^2 \eta\\
& +\left[ i \frac{\omega}{v} \biggl(1+\chi \gamma^2 \frac{v^2}{c^2} \biggr) + \frac{1}{2}  \gamma^2 \frac{v^2}{c^2}  (\partial_x \chi) 
+\frac{1}{2\chi}  (\partial_x \chi) +O(\epsilon^2) \right] \eta'+(\cdots+O(\epsilon^2)) \eta=0,
\end{split}
\eeq
where we have not written explicitly  the $O(\epsilon^2)$ terms and the last contribution because they are not useful herein. 
In particular, the last contribution does not affect the indicial equation for the reduced equation
\beq
-\biggl(1- \chi \gamma^2 \frac{v^2}{c^2}\biggr) \partial_x^2 \eta 
+\left[ i \frac{\omega}{v} \biggl(1+\chi \gamma^2 \frac{v^2}{c^2} \biggr) + \frac{1}{2}  \gamma^2 \frac{v^2}{c^2}  (\partial_x \chi) 
+\frac{1}{2\chi}  (\partial_x \chi) \right] \eta'+(\cdots) \eta=0.
\eeq
We find 
\beq
\alpha_1 =0, \qquad \alpha_2 = -i \frac{\omega c}{\gamma^2 v^2 n'} ,
\eeq 
from which
\beq
\lambda=1 +i \frac{\omega c}{\gamma^2 v^2 n'}. 
\eeq

\subsubsection{WKB approximation}
 
In this case we put 
\beq
\eta (x) = \exp( \frac{\theta(x)}{\epsilon} ) \sum_{i=0}^\infty \epsilon^i y_i (x) ,
\eeq
and obtain again
\beq
\label{eik-di}
\theta^{'4} + \frac{1}{\chi \gamma^2 v^2}\biggl(1- \chi \gamma^2 \frac{v^2}{c^2}\biggr) \theta^{' 2} =0.
\eeq
By now, we consider just the case $x>0$, as the case $x<0$ is analogous to the one of the Corley model. 
Coming back to \eqref{eik-di}, its solutions are $\theta'=0$ (multiplicity two), and 
\beq
\theta'_\pm = \pm i \frac{1}{\sqrt{\chi} \gamma v}\sqrt{1-\chi \gamma^2 \frac{v^2}{c^2} }.
\eeq
The latter solutions are associated with the transport equation
\beq
y'_0+\frac{1}{(1-\chi \gamma^2 \frac{v^2}{c^2} )} \biggl[ -\frac{3}{4} \frac{\chi'}{\chi} +
i \frac{\omega}{2 v} \biggl(1+\chi \gamma^2 \frac{v^2}{c^2} \biggr) 
-\frac{1}{4} \biggl(1-\chi \gamma^2 \frac{v^2}{c^2} \biggr) \frac{\chi'}{\chi}\biggr] y_0=0.
\eeq
Solutions are of the form 
\beq
y_0 (x) = A \, \chi (1-a)^{-\frac{3}{4}} e^{-i \frac{\omega}{2 v} \int^x ds 
\frac{1+a(s)}{1-a(s)}} ,
\eeq
and the high momentum modes are
\beq
\eta_\pm (x) = e^{\pm \frac{i}{\epsilon} \frac{1}{\gamma v} \int^x ds \frac{1}{\sqrt{\chi (s)}} 
\sqrt{1-a(s) }} y_0 (x).
\eeq
Near the TP one obtains 
\beq
|\eta_\pm (x)| \propto x^{-3/4}.
\eeq
Two further solutions occurring when $\theta'=0$ are obtained from the reduced equation.
We find near the regular singular point $x=0$ (our TP) the series expansions
\begin{align}
\eta_{-s} (x) &= 1+\sum_{n=1}^{\infty} c_n x^n,\\
\eta_{+s} (x) &= x^{-i \frac{\omega c}{\gamma^2 v^2 n'}}  \biggl( 1+\sum_{n=1}^{\infty} d_n x^n\biggr) .
\end{align}
Near the TP we get 
\begin{align}
|\eta_{-s} (x)| &\propto \text{const},\\
|\eta_{+s} (x)| &\propto \text{const}, 
\end{align}
and all the aforementioned asymptotics have  the same behaviour as calculated in~\cite{hopfield-kerr} for the electromagnetic case and for the simpler $\phi$--$\psi$ model in~\cite{hawbook}, chapter 10.

\subsubsection{Approximation near the turning point}

We recall that $\lambda=1 +i \frac{\omega c}{\gamma^2 v^2 n'} \coloneqq 1 -i \frac{\omega c}{\kappa}$;
being $z=(\frac{2 \kappa}{v c^3})^{1/3} \epsilon^{-2/3} x$, 
we find in the external region $x>0$
\begin{align}
\label{saddle-p-phi}
w_+ (z) &\simeq \frac{1}{2 \sqrt{\pi}} e^{-\frac{3}{4}\pi i} e^{\frac{\pi \omega c}{2 \kappa}} |z|^{-\frac{i\omega c}{2\kappa}-\frac{3}{4}} e^{i\frac{2}{3} |z|^{3/2}}, \\
\label{saddle-m-phi}
w_- (z) &\simeq \frac{1}{2 \sqrt{\pi}} e^{\frac{1}{4}\pi i} e^{-\frac{\pi \omega c}{2 \kappa}} |z|^{-\frac{i\omega c}{2\kappa}-\frac{3}{4}} e^{-i\frac{2}{3} |z|^{3/2}}.
\end{align}
As to the cut contribution, we find 
\beq
w_{cut} (z) \simeq - \frac{1}{i \pi} \Gamma \Bigl(-i\frac{\omega c}{\kappa}\Bigr) \sinh (\frac{\pi \omega c}{\kappa}) |z|^{i \frac{\omega c}{\kappa}}. 
\label{cut-die-phi}
\eeq
Also in this case, we obtain 
for $x<0$ the decaying mode in an analogous way as for~\eqref{decay-sub}. {\color{black} It is worthwhile noting that, due to the universal form 
of the equation \eqref{four-z} governing the near-horizon approximation, the approximate expressions for the aforementioned modes near horizon are of the 
same type as for the simpler Corley model \cite{corley,cpf,Coutant-thick}, as the latter is a subcase of the general framework we are discussing.} 

Near the turning point we obtain from the black hole boundary condition and in the external region 
\beq
\label{nearhorizon}
\phi (x,t) = \phi_1 (x,t) + \phi_2 (x,t) + \phi_3 (x,t)+h \phi_4 (x,t), 
\eeq
where $\phi_1 \mapsto w_+$, $\phi_2 \mapsto w_-$, $\phi_3 \mapsto w_{cut}$ and $\phi_4 \mapsto 1$.  {\color{black} As far as the factor $h$ 
is concerned, analogous considerations as in the case of the previous section hold true.} 
By comparing with the WKB solutions 
again in the matching region, we find
\beq
\begin{split}
\label{connection-phi}
\phi (x,t) &= \frac{1}{2\sqrt{\pi}} e^{{\frac{\omega c}{2\kappa} \pi }} e^{-i \frac{3}{4}\pi} \frac{v^2 \gamma^2}{c^2} \sqrt{\frac{2\kappa}{v}}
\epsilon^{i \frac{\omega c}{3\kappa}+\frac{1}{2}} \biggl(\frac{2\kappa}{v c^3}\biggr)^{-i \frac{\omega c}{6\kappa}}\varphi_+ (x,t)\\ 
& \quad + \frac{1}{2\sqrt{\pi}} 
e^{-{\frac{\omega c}{2\kappa} \pi }} e^{i \frac{1}{4}\pi} \frac{v^2 \gamma^2}{c^2} \sqrt{\frac{2\kappa}{v}}
\epsilon^{i \frac{\omega c}{3\kappa}+\frac{1}{2}} \biggl(\frac{2\kappa}{v c^3}\biggr)^{-i \frac{\omega c}{6\kappa}} \varphi_- (x,t) \\
& \quad -
\frac{\sinh (\frac{\omega c}{ \kappa})}{\pi i} \Gamma \Bigl(- i \frac{\omega c}{ \kappa}\Bigr)  \biggl(\frac{2\kappa}{v c^3}\biggr)^{i \frac{\omega c}{\kappa}} 
\epsilon^{-i \frac{2\omega c}{3\kappa}} 
\varphi_{+s} (x,t)+ h \varphi_{-s} (x,t). 
\end{split}
\eeq
A trivial matching involves also the fourth mode $\varphi_{-s} (x,t)$, which is regular everywhere.

\subsection{Thermality}

We can identify the aforementioned solutions as corresponding to 
the backward state $B\mapsto \phi_{-s} (x,t)$, the positive high-momentum state $P\mapsto \phi_{+} (x,t)$, the 
negative norm high-momentum state $N\mapsto\phi_{-} (x,t)$ and the Hawking state $H\mapsto \phi_{+s} (x,t)$, respectively.  
We obtain 
\beq
\label{thermal}
\frac{|N|^2}{|P|^2} \coloneqq \frac{|J_x^{-}|}{|J_x^{+}|}= e^{-\frac{2 \pi c}{\kappa} \omega},
\eeq
which corresponds to the standard signal of the thermal character of the black hole horizon. 
The current density has the following structure~\cite{hopfield-hawking}: 
\beq
\label{curr-j}
J^\mu \coloneqq \frac{i}{2} \biggl[ \phi^* \partial^\mu \phi - (\partial^\mu \phi^*) \phi 
+ \frac{1}{\chi \omega_0^2} v^\mu \psi^* v^\alpha \partial_\alpha \psi - 
\frac{1}{\chi \omega_0^2} v^\mu \psi v^\alpha \partial_\alpha \psi^* +\frac{1}{c} v^\mu (\psi^* \phi - \psi \phi^*)\biggr] .
\eeq
One considers the fields in the asymptotic (homogeneous) region in the comoving frame, where 
they are normalized as in~\cite{phi-psi}. Furthermore, 
the term quadratic in $\psi$ in the present expansion at the leading order is suppressed, as is $O(\epsilon^2)$. 
One obtains
\beq
|J_x|= \abs{ \Biggl(-k_x -\frac{1}{c^2} \chi \gamma v \frac{k_\alpha v^\alpha}{1-\frac{(k_\alpha v^\alpha)^2}{\omega_0^2}} +O(\epsilon^2)\Biggr)} \quad
\abs{\Bigl(-k_x -\frac{1}{c^2} \chi \gamma v (k_\alpha v^\alpha)\Bigr) \varphi^\ast \varphi}.
\eeq
In particular, in the asymptotic region $x\to \infty$ we have 
\begin{align}
k_x^{+s} &= \frac{\omega}{v} \frac{n-\frac{v}{c}}{\frac{c}{v}-n},\\
k_x^{-s} &= -\frac{\omega}{v} \frac{n+\frac{v}{c}}{\frac{c}{v}+n}.
\end{align}

\subsection{The grey-body factor}

Of course, one may study the problem of determining $h$ directly by considering the reduced equation and its solutions. This might be a nontrivial route, 
as the equation is quite involved.
Alternatively, in order to calculate the grey-body factor at least in an approximate way, we could first identify the metric associated with the model at hand. 
From equations~\eqref{phi-eq},~\eqref{psi-eq}, in the approximation where the term $\propto \epsilon^2$ is 
neglected and in the eikonal approximation, we get the metric also deduced in~\cite{belgiorno-prd}
\beq 
{ds}^2 = c^2 \gamma^2 \frac{1}{n^2} \biggl(1+\frac{n v}{c}\biggr)\biggl(1-\frac{n v}{c}\biggr) {dt}^2
+ 2 \gamma^2 \frac{v}{n^2} \bigl(1-n^2\bigr) dt dx -\gamma^2 \biggl(1+\frac{v}{n c}\biggr)\biggl(1-\frac{v}{n c}\biggr) {dx}^2,
\eeq 
where we are in the comoving frame of the pulse generating a propagating dielectric perturbation and 
the refractive index depends on $x$: $n=n(x)$. 
Differently from the Corley model, the metric is not exact but approximated, and holds only in the eikonal approximation. 
The above metric is conformally related to the one deduced in~\cite{linder}. There exists a coordinate transformation carrying the metric
into a static form; even if they are singular, we 
carry out the relative transformation because it allows a 
direct computation of the grey-body coefficient.
The following coordinate change
\beq
dt = d\tau -\alpha (x) dx,
\eeq
where
\beq
\alpha (x) = \frac{g_{01} (x)}{g_{00} (x)}.
\eeq
carries the metric to the static form~\cite{belgiorno-prd}
\beq\label{static-pulse}
{ds}^2 = \frac{c^2}{n^2 (x)} g_{\tau \tau} (x) {d\tau}^2 - \frac{1}{g_{\tau \tau} (x)} {dx}^2,
\eeq
where
\beq\label{pulse-lapse}
g_{\tau \tau} (x) \coloneqq \gamma^2 \biggl(1+n(x)\frac{v}{c}\biggr)\biggl(1-n(x) \frac{v}{c}\biggr) .
\eeq
We do not delve into the explicit calculation, as is the same displayed in \cite{belgiorno-prd}, which confirms in the present two-dimensional model that 
$\Gamma = 1$, and then, in this approximation, $h=0$ once more, and that there is a divergence as $\omega\to 0$ in the number of created particles, as numerically tested in~\cite{finazzi-carusotto} and then also found in~\cite{linder} 
in a different approximation scheme (see also~\cite{petev}). This approximation might be too crude, and $\Gamma<1$ could also be allowed 
by a better approximation. Still, again, the leading contribution to $h$ as arising from the pair creation process  is  vanishing.

Also in this case, a maximal frequency $\omega_{max}$ exists~\cite{finazzi-carusotto} (see also~\cite{hopfield-hawking}) beyond which 
no Hawking effect is expected, and then a truncation of the spectrum for $\omega>\omega_{max}$ is to be taken into account.  
One may wonder which differences occur with respect to the calculation in~\cite{linder}. Therein, 
the fourth backward mode participates to the Corley's diagram near the TP, as it appears as a 
further cut integral in the Fourier space. It is remarkable that this diagram was calculated 
in the approximation where the square of the resonance frequency is a linear function in $x$, which is of course different from 
the case at hand. But this is not the only source of differences, as it is the approximation we perform herein 
in itself which  is able to leave just only a cut-integral (in the Laplace dual space), with the other short wavenumber 
mode (the backward one) absent from the diagram. Analogous considerations 
can be made in a comparison with the calculations developed for the Hopfield model discussed in~\cite{hopfield-kerr}, where 
the grey-body factor was not available.

\section{Conclusions}

We have explored a further way to approach analytical calculations for the Hawking effect in 
analogue gravity. A fourth order equation, which is of the Orr--Sommerfeld type, has been shown to play the 
role of master equation in analogue gravity, with reference to the analogous Hawking effect. 
The approximation adopted is the one of weak dispersive effects, where the 
suitable coupling of the fourth order term is associated with the parameter $\epsilon$ entering the equation.  {\color{black} 
This kind of approximation is not new in literature, see e.g. \cite{cpf,Coutant-thick}, but it is applied in the framework 
provided by Nishimoto's analysis~\cite{ni} for equations of the Orr--Sommerfeld.  This allows us to  
achieve}   a suitable approximation near the turning point (horizon), and we are enabled to provide a complete study of thermality for both the subluminal fluid model of \cite{corley-jacobson,corley} and for the dielectric one. Indeed, we can provide a scheme for the calculation of an analytic expression of the grey-body factor, which is {\color{black} in agreement with the analysis carried out in \cite{cpf,Coutant-thick}, as far as the Corley's model is concerned, 
but is more general and allows to encompass important physical models which cannot be included by the Corley's model itself: dielectrics, 
BEC and water waves with varying speed of sound velocity $c(x)$. Indeed, the same calculational scheme 
can be adopted successfully also in the case of BEC and of surface waves in the companion paper~\cite{belbecsw}. }

Then a more complete study of the Hawking 
emission in condensed matter systems is achieved when dispersion is weak, which provides the most direct correspondence with the 
standard Hawking effect, with an enhanced role of the reduced equation (i.e. the equation one obtains in absence of dispersion).\\ 
It is remarkable that the geometrical setting of the analogous Hawking effect in this scheme arises in the WKB approximation 
which holds near but not too near the horizon. 
The model of course leaves open the possibility to explore more sophisticated situations where 
dispersive effects are strong, which would provide regimes for Hawking-like radiation which are more far from the standard case. 

The perspective is open also for a more sophisticated analogue black hole spectroscopy, allowing a more 
precise comparison between experimental measurements and theoretical computations.

\section*{Acknowledgements}

F.B. thanks Dario Pierotti for some discussions concerning mathematical aspects of the paper.
A.V.~was partially supported by MIUR-PRIN contract 2017CC72MK\_003.

\appendix
\section{A further justification of the near horizon approximation}

We provide a further justification of the near horizon approximation, which allows us also to show that 
the Orr--Sommerfeld form of the equation is not mandatory, in the sense that one can allow also for third 
order terms in the derivative, with the only restriction that they are at least of the same order of the fourth order one 
in the suitable coupling and that they do not vanish at the TP.

We start by a slight generalization of~\eqref{orso}
\beq
\delta^2 \frac{d^4 \Phi}{dx^4}\pm\left[ \delta^2 p_4 (x,\delta) \frac{d^3 \Phi}{dx^3} + p_3 (x,\delta) \frac{d^2 \Phi}{dx^2}+
p_2 (x,\delta) \frac{d \Phi}{dx} +p_1 (x,\delta) \Phi \right]=0, 
\label{orso-eta}
\eeq
where we have changed the power of the expansion parameter  with respect to~\cite{reid-74,drazin}, in order to 
allow a direct comparison with the framework discussed in the paper.
The new term in the third order derivative has been added. We first introduce for simplicity of notation 
\beq
f(x) \coloneqq \frac{p_{30} (x)}{p'_{30}(0)},
\label{langer-f}
\eeq
where we have shifted the turning point (where  $p_{30} (x)=0$) at $x=0$.

Then we define a Langer-like variable, adapting the definition assumed in~\cite{reid-74,drazin}: 
\beq
\eta (x) \coloneqq \left[ \frac{3}{2} \int_0^x dy \sqrt{f(y)} \right]^{2/3}.
\eeq
For definiteness, we consider the subluminal case (the superluminal one is obtained in a straightforward way). 
We shall indicate with $\Phi^{(i)}$, $i=1,2,3,4$ the derivatives with respect to the new variable, and by 
$\Phi'$, $\Phi''$, $\Phi'''$, $\Phi''''$ the derivatives with respect to $x$.
We can notice that 
\beq
\eta' (x)= \sqrt{\frac{f(x)}{\eta(x)}},
\eeq
which is regular as $x\to 0$. Furthermore, due to~\eqref{langer-f}, also $\eta' \to 1$ as $x\to 0$ holds true. 

As to~\eqref{orso-eta}, considering only the leading order terms, we obtain
\beq
\label{orso-delta}
\begin{split}
&\delta^2 \Phi^{(4)} +\delta^2 \left( 6 \frac{\eta''}{(\eta')^2} + p_{40}  \frac{1}{\eta'} +O(\delta)\right) \Phi^{(3)} + 
\left( p'_{30} (0) \eta +O(\delta)\right)  \Phi^{(2)} \\
&+\left( p_{20}  \frac{1}{(\eta')^3} + p_{30}  \frac{\eta''}{(\eta')^4} +O(\delta) \right) \Phi^{(1)} +
\left(p_{10} \frac{1}{(\eta')^4} + O (\delta) \right) \Phi=0. 
\end{split}
\eeq
We now define 
\beq
\epsilon_R \coloneqq \frac{\delta}{(p'_{30} (0))^{1/2}}, 
\eeq
in order to mimic the behavior occurring in~\cite{reid-74,drazin}. An equation holding in the near horizon 
approximation is obtained by means of the following definition (with some abuse of notation) 
\beq
\Phi (\zeta, \epsilon_R) \coloneqq \Phi \biggl(\frac{\eta}{(\epsilon_R)^{2/3}}, \epsilon_R\biggr),
\eeq
and also the new variable 
\beq
\zeta \coloneqq  \frac{\eta}{(\epsilon_R)^{2/3}}=(p'_{30} (0))^{1/3} \delta^{-2/3} \eta. 
\eeq
Furthermore, one has to take into account that 
\beq
p_{30} = (\eta')^2 p'_{30}(0) \eta.
\eeq
Then one finds the following equation
\beq
\label{orso-eps}
\begin{split}
&(p'_{30} (0))^{4/3} \Phi^{(4)} + \delta^{2/3} \left( 6 \frac{\eta''}{(\eta')^2} + p_{40}  \frac{1}{\eta'} +O(\delta)\right) (p'_{30} (0)) \Phi^{(3)} + 
\left( (p'_{30} (0))^{4/3}  \zeta +O(\delta^{1/3} )\right)  \Phi^{(2)} \\
&+\left[ \left( p_{20} (p'_{30} (0))^{1/3} \Phi^{(4)}  \frac{1}{(\eta')^3} \right) +O(\delta^{2/3} ) \right] \Phi^{(1)} +
O (\delta^{2/3} ) \Phi=0. 
\end{split}
\eeq
At the leading order and assuming that $p_{2}$ (and then also $p_{20}$) is analytic in a neighbourhood of the TP we obtain 
\beq
\Phi^{(4)} + \zeta  \Phi^{(2)}+\frac{ p_{20}(0)}{p'_{30} (0)}\Phi^{(1)} =0,
\label{nher}
\eeq
which by taking into account that 
\beq
\lambda \coloneqq \frac{ p_{20}(0)}{p'_{30} (0)},
\eeq
coincides with the equation obtained by means of the method borrowed from~\cite{ni}. The relation between 
$\epsilon$ in the previous sections and $\epsilon_R$ is simply 
\beq
\epsilon = \epsilon_R \sqrt{p'_{30} (0)}=\delta,
\eeq
and also $\zeta=z$ holds.

\section{matching conditions}
\label{matching}

Let us now further discuss the matching conditions underlying the scattering process at hand. 
We take into consideration states living on the right side of the turning point, directly involved in the Hawking effect.  
We have to match in a single solution the WKB part and the near horizon part of the modes introduced above, in such a way to 
obtain basis functions which are defined in the whole domain. 
For the WKB part, we have to consider the basis
\beq
\{ \varphi_+^{WKB} (x),  \varphi_-^{WKB} (x), \varphi_{+s}^{WKB} (x),\varphi_{-s}^{WKB} (x) \},
\eeq
whereas for the near horizon (NH)  region we get the further basis 
\beq
\{ \varphi_+^{NH} (x),  \varphi_-^{NH} (x), \varphi_{+s}^{NH} (x),\varphi_{-s}^{NH} (x) \}.
\eeq
Let us denote $\varphi_i^{WKB} (x)$ and $\varphi_i^{NH} (x)$ the parts to be joined for the $i$-mode, with 
$i=\pm,\pm s$. The general WKB solution has the form 
\beq
\varphi^{WKB} (x)= \sum_i C_i \varphi_i^{WKB} (x),
\eeq
where $C_i$ are constant (i.e. independent from $x$), and the general NH solution is 
\beq
\varphi^{NH} (x)= \sum_i D_i \varphi_i^{NH} (x),
\eeq
where also $D_i$ are constant. In the matching region, where the two approximation co-exist, we have 
\beq
\varphi_i^{WKB} (x)\sim a_i h_i (x),
\eeq
and also
\beq
\varphi_i^{NH} (x)\sim b_i h_i (x),
\eeq
with the same functional dependence $h_i (x)$. Then, matching in the linear region requires 
\beq
C_i = \frac{b_i}{a_i} D_i.
\eeq
Notice that, compared to the standard matching of the WKB solutions with Airy functions for the Schr\"odinger equation 
in quantum mechanics, in place of fixing the constant for the near turning point solutions as functions of the ones in the 
WKB-allowed regions, in agreement with Corley's ideas, we proceed in the complementary direction, as an indication that 
part of the amplitudes arises from what happens at the turning point.\\
Moreover, for $x\to \infty$ the propagating modes participating to the Hawking process behave as plane waves: 
\beq
\varphi_i^{WKB} (x)\sim \bar{a}_i e^{i k_i (\omega) x},
\eeq
so that 
\beq
\varphi^{WKB} (x)= \sum_i  C_i \bar{a}_i e^{i k_i (\omega) x},
\eeq
and we may define  
\beq
c_i \coloneqq \frac{b_i}{a_i} D_i \bar{a}_i,  
\eeq
in order to compare with the amplitudes defined in \cite{corley}. In order to get scattering amplitudes, let us 
write 
\beq
C_i = \bar{C}_i N_i,
\eeq
where $N_i$ are the normalizations of the modes in the asymptotic region, which are consistent with the 
quantization of the field in the $\omega$-representation \cite{macher-bec}. In particular, we have
\beq
N_i = \frac{1}{\sqrt{4\pi |v_g (k_i (\omega) ) (\omega -v k_i (\omega)|}},
\eeq
where $v_g (k_i (\omega) )$ is the group velocity of the $i-th$ mode. 
$\bar{C}_j$ represent the 
actual amplitudes: 
\beq
\bar{C}_i = \frac{b_i}{N_i a_i} D_i \bar{a}_i.
\eeq
Because of the black hole boundary condition, we have 
\beq
D_+=D_-=D_{+s}\coloneqq D,
\eeq
whereas the fourth mode has a different amplitude, that we put equal to 
\beq
D_{-s} \eqqcolon h D.
\eeq
Therefore, by comparison with the asymptotic behavior of the field, one obtains 
\beqnl
\label{amplitudes} 
&& \bar{C}_+= \frac{b_+ \bar{a}_+}{N_+ a_+} \frac{N_{+s} a_{+s}}{b_{+s} \bar{a}_{+s}},\\
&& \bar{C}_-= \frac{b_- \bar{a}_-}{N_- a_-} \frac{N_{+s} a_{+s}}{b_{+s} \bar{a}_{+s}}, \label{cimeno}\\
&& \bar{C}_{+s}=1,\\
&& \bar{C}_{-s}= h \frac{b_{-s} \bar{a}_{-s}}{N_{-s} a_{-s}} \frac{N_{+s} a_{+s}}{b_{+s} \bar{a}_{+s}}.
\eeqnl
See also the comment below equation \eqref{currents}. 
As regards the complete solution, we have a basis 
\beq
\{ \varphi_+(x),  \varphi_-(x), \varphi_{+s}(x),\varphi_{-s} (x) \},
\eeq
which reduces to the aforementioned bases in the different regions: of course 
\beq
\varphi_i (x) \sim \varphi_i^{WKB}(x)
\eeq
asymptotically, and also 
\beq
\varphi_i (x) \sim \varphi_i^{NH}(x)
\eeq
near the turning point. In the matching region it holds
\beq
\varphi_i (x) \sim \bar{C}_i N_i a_i h_i (x) = D_i b_i h_i (x).
\eeq
For the process at hand, the general solution
\beq
\varphi (x) =\sum_i A_i \varphi_i (x)
\eeq
must be such that, 
asymptotically, one gets again 
\beq
A_i = 
\bar{C}_i = \frac{b_i}{N_i a_i} D_i \bar{a}_i.
\eeq
It is worthwhile mentioning that, in more rigorous mathematical terms, we have been discussing the 
topic of central connections in terms of the language adopted in \cite{ni}. Therein, one considers a fundamental 
matrix $\Phi^{NH}$ of solutions near the TP, a fundamental matrix $\Phi^{WKB}$ of solutions in the WKB region, and then 
matches through a matrix $\Lambda$ according to 
\beq
\Phi^{WKB}=\Phi^{NH} \Lambda,
\eeq
where $\Lambda$ is asymptotically diagonal \cite{ni}. It is easily verified that this condition is 
equivalent to the one we discussed above.




\begin{thebibliography}{99}

\bibitem{brout}
R.Brout, S.Massar, R.Parentani and Ph. Spindel, 
Phys. Rev. D, 52, 4559 (1995).

\bibitem{corley}
S. Corley, 
Phys. Rev. D, 57, 6280 (1998). 
[hep-th/9710075]. 

\bibitem{himemoto}
Y. Himemoto,  and T. Tanaka, 
Phys. Rev. D, 61, 064004 (2000). 
[gr-qc/9904076].

\bibitem{saida}
H. Saida, and M. Sakagami, 
Phys. Rev. D, 61, 084023 (2000).  [gr-qc/9905034].


\bibitem{Schutzhold-Unruh} 
  R.~Schutzhold and W.~G.~Unruh,
  Phys.\ Rev.\ D {\bf 78}, 041504 (2008)
  [arXiv:0804.1686 [gr-qc]].

\bibitem{unruh-s}

W.G. Unruh, and  R. Sch\"{u}tzhold, 
Phys. Rev. D 71,
024028  (2005).  [gr-qc/0408009].

\bibitem{balbinot} 
  R.~Balbinot, A.~Fabbri, S.~Fagnocchi and R.~Parentani,
  Riv.\ Nuovo Cim.\  {\bf 28}, 1 (2005)
  [gr-qc/0601079].


\bibitem{cpf} 
  A.~Coutant, R.~Parentani and S.~Finazzi,
  Phys.\ Rev.\ D {\bf 85}, 024021 (2012)
  [arXiv:1108.1821 [hep-th]].

\bibitem{Leonhardt-Robertson} 
  U.~Leonhardt and S.~Robertson,
  New J.\ Phys.\  {\bf 14}, 053003 (2012).

\bibitem{Coutant-prd} 
  A.~Coutant, A.~Fabbri, R.~Parentani, R.~Balbinot and P.~Anderson,
  Phys.\ Rev.\ D {\bf 86}, 064022 (2012)
  [arXiv:1206.2658 [gr-qc]].

\bibitem{Coutant-und} 
  A.~Coutant and R.~Parentani,
   Phys.\ Fluids {\bf 26}, 044106 (2014)
  arXiv:1211.2001 [physics.flu-dyn].

\bibitem{Un-schu}
  R.~Schutzhold and W.~G.~Unruh,
  Phys.\ Rev.\ D {\bf 88}, 124009 (2013)
  arXiv:1308.2159 [gr-qc].

\bibitem{petev} 
  M.~Petev, N.~Westerberg, D.~Moss, E.~Rubino, C.~Rimoldi, S.~L.~Cacciatori, F.~Belgiorno and D.~Faccio,
  Phys.\ Rev.\ Lett.\  {\bf 111}, 043902 (2013)
  doi:10.1103/PhysRevLett.111.043902
  [arXiv:1303.5967 [physics.optics]].

\bibitem{Coutant-thick} 
  
A.~Coutant and R.~Parentani,
  Phys.\ Rev.\ D {\bf 90}, no. 12, 121501 (2014)
  [arXiv:1402.2514 [gr-qc]].
 
\bibitem{hopfield-hawking} 
  F.~Belgiorno, S.~L.~Cacciatori and F.~Dalla Piazza,
  Phys.\ Rev.\ D {\bf 91}, no. 12, 124063 (2015)
  doi:10.1103/PhysRevD.91.124063
  [arXiv:1411.7870 [gr-qc]].

\bibitem{linder} 
  M.~F.~Linder, R.~Schutzhold and W.~G.~Unruh,
  Phys.\ Rev.\ D {\bf 93}, no. 10, 104010 (2016)
  doi:10.1103/PhysRevD.93.104010
  [arXiv:1511.03900 [gr-qc]].

\bibitem{philbin-exact} 
  T.~G.~Philbin,
  Phys.\ Rev.\ D {\bf 94}, no. 6, 064053 (2016)
  doi:10.1103/PhysRevD.94.064053
  [arXiv:1607.03743 [gr-qc]].

\bibitem{hopfield-kerr} 
  F.~Belgiorno, S.~L.~Cacciatori, F.~Dalla Piazza and M.~Doronzo,
  Phys.\ Rev.\ D {\bf 96}, no. 9, 096024 (2017)
  doi:10.1103/PhysRevD.96.096024
  [arXiv:1707.01663 [hep-th]].

\bibitem{coutant-subcritical} 
  A.~Coutant and S.~Weinfurtner,
  Phys.\ Rev.\ D {\bf 94}, no. 6, 064026 (2016)
  doi:10.1103/PhysRevD.94.064026
  [arXiv:1603.02746 [gr-qc]].

\bibitem{coutant-kdv} 
  A.~Coutant and S.~Weinfurtner,
  Phys.\ Rev.\ D {\bf 97}, no. 2, 025005 (2018)
  doi:10.1103/PhysRevD.97.025005
  [arXiv:1707.09651 [gr-qc]].

\bibitem{coutant-bdg} 
  A.~Coutant and S.~Weinfurtner,
  Phys.\ Rev.\ D {\bf 97}, no. 2, 025006 (2018)
  doi:10.1103/PhysRevD.97.025006
  [arXiv:1707.09664 [gr-qc]].


\bibitem{corley-jacobson}
S. Corley and T. Jacobson, 
Phys. Rev. D, 54, 1568 (1996). 

\bibitem{belbecsw}
F.~Belgiorno, S.~L.~Cacciatori, A.~Farahat, and A.~Vigan\`o, Analogue Hawking Effect: BEC and Surface Waves. To appear (2019). 



\bibitem{la}
R.E. Langer, Formal Solutions and a Related Equation For a Class of Fourth Order Differential Equations of a Hydrodynamic Type. Transactions of the American Mathematical Society, vol. 92, no. 3, 1959, pp. 371-410. 

\bibitem{ni}

T.Nishimoto, K\u{o}dai Math. Sem. Rep. 29, (1978), 233.

\bibitem{ni-I}

T.Nishimoto, K\u{o}dai Math. Sem. Rep. 24, (1972), 281.

\bibitem{ni-global}

T.Nishimoto, K\u{o}dai Math. Sem. Rep. 27, (1976), 128.

\bibitem{ni-tp}

T.Nishimoto, K\u{o}dai Math. Sem. Rep. 20, (1968), 218.


\bibitem{kanwal} 
Ram P.Kanwal, Generalized Functions. Theory and Applications. Third edition. Birkh\"{a}user, Boston (2004). 


\bibitem{holmes}

M.H.Holmes, Introduction to Perturbation Methods. Texts in Applied Mathematics, Vol. 20. Springer, Berlin (2013).

\bibitem{eastham}

M.S.P. Eastham, The Asymptotic Solution of Linear Differential Systems: Application of the Levinson Theorem. 
London Mathematical Society Monographs New Series, Vol. 4. Clarendon Press, 1989.


\bibitem{olver}
Frank W.I. Olver, {\it Asymptotics and Special Functions}. CRC Press, New York (1997). 

\bibitem{wong}
R.Wong, {\it Asymptotic Approximation of Integrals}. Academic Press, New York (1989). 

\bibitem{liberati} 
  A.~Coutant, S.~Finazzi, S.~Liberati and R.~Parentani,
  Phys.\ Rev.\ D {\bf 85}, 064020 (2012)
  doi:10.1103/PhysRevD.85.064020
  [arXiv:1111.4356 [gr-qc]].




\bibitem{miller}
P.D.Miller, {\it Applied Asymptotic Analysis}. Graduate Studies in Mathematics, Volume 75. American Mathematical Society, 
Providence, Rhode Island (2006).

\bibitem{visser-ess}
   Matt Visser, 
  Int.\ J.\ Mod.\ Phys.\ D {\bf 12}, 649 (2003)
  doi:10.1142/S0218271803003190
  [hep-th/0106111].

\bibitem{anderson-exact} 
  A.~Fabbri, R.~Balbinot and P.~R.~Anderson,
  Phys.\ Rev.\ D {\bf 93}, no. 6, 064046 (2016)
  doi:10.1103/PhysRevD.93.064046
  [arXiv:1512.08447 [gr-qc]].

\bibitem{macher-wh} 
  J.~Macher and R.~Parentani,
  Phys.\ Rev.\ D {\bf 79}, 124008 (2009)
  doi:10.1103/PhysRevD.79.124008
  [arXiv:0903.2224 [hep-th]].


\bibitem{macher-bec} 
  J.~Macher and R.~Parentani,
  Phys.\ Rev.\ A {\bf 80}, 043601 (2009)
  doi:10.1103/PhysRevA.80.043601
  [arXiv:0905.3634 [cond-mat.quant-gas]].



\bibitem{hawbook} 
  F.~D.~Belgiorno, S.~L.~Cacciatori and D.~Faccio,
  ``Hawking Radiation : From Astrophysical Black Holes to Analogous Systems in Lab,''
  World Scientific Publishing Company, Singapore (2018). ISBN 9789814508537.

\bibitem{phi-psi} 
  F.~Belgiorno, S.~L.~Cacciatori, F.~Dalla Piazza and M.~Doronzo,
  Eur.\ Phys.\ J.\ C {\bf 76}, no. 6, 308 (2016)
  doi:10.1140/epjc/s10052-016-4146-1
  [arXiv:1512.08738 [math-ph]].

\bibitem{belgiorno-prd} 
  F.~Belgiorno, S.~L.~Cacciatori, G.~Ortenzi, L.~Rizzi, V.~Gorini and D.~Faccio,
  Phys.\ Rev.\ D {\bf 83}, 024015 (2011)
  doi:10.1103/PhysRevD.83.024015
  [arXiv:1003.4150 [quant-ph]].

\bibitem{finazzi-carusotto} 
  S.~Finazzi and I.~Carusotto,
  Phys.\ Rev.\ A {\bf 89}, no. 5, 053807 (2014)
  doi:10.1103/PhysRevA.89.053807
  [arXiv:1303.4990 [physics.optics]].

\bibitem{reid-74}
W.H.Reid, 
Studies in Appl. Math. {\bf 53}, 217 (1974).

\bibitem{drazin}
P.G.Drazin and W.H.Reid, Hydrodynamic Stability. Cambridge Mathematical Library, Cambridge University Press, 
Cambridge (2004).



\end{thebibliography}
\end{document}